\documentclass[pre,twocolumn]{revtex4-1}

\usepackage{amssymb}
\usepackage{amsmath}
\usepackage{amsfonts}
\usepackage{graphicx}

\begin{document}

\title{Geophysical implications of a decentered inner core}
\author{C\u{a}lin Vamo\c{s}}
\email{cvamos@ictp.acad.ro}
\affiliation{ ``T. Popoviciu'' Institute
of Numerical Analysis, Romanian Academy, P.O. Box 68, 400110
Cluj-Napoca, Romania}
\author{Nicolae Suciu}
\email{suciu@math.fau.de}
\affiliation{Department of Mathematics, Friedrich-Alexander University
of Erlangen-Nuremberg, Cauerstr. 11, 91058 Erlangen, Germany}

\begin{abstract}
In a first approximation, the Earth's interior has an isotropic
structure with a spherical symmetry. Over the last decades the
geophysical observations have revealed, at different spatial scales,
the existence of several perturbations from this basic structure. In
this paper we discuss the hemispheric perturbations induced to this
basic structure if the inner core is displaced from the center of
mass of the Earth. Using numerical simulations of the observed
hemispheric asymmetry of the seismic waves traveling through the
upper inner core, with faster arrival times and higher attenuation
in the Eastern Hemisphere, we estimate that the present position of
the inner core is shifted by tens of kilometers from the Earth's
center eastward in the equatorial plane. If the only forces acting
on the inner core were the gravitational forces, then its
equilibrium position would be at the Earth's center and the
estimated displacement would not be possible. We conjecture that,
due to interactions with the flow and the magnetic field inside the
outer core, the inner core is in a permanent chaotic motion. To
support this hypothesis we analyze more than ten different
geophysical phenomena consistent with an inner core motion dominated
by time scales from hundreds to thousands of years.
\end{abstract}

\maketitle

\section{Introduction}

Situated at the Earth's center, surrounded by the fluid outer core
(OC), the inner core (IC) is primarily composed of a solid iron and
nickel alloy. For a long time after its discovery by Lehmann in 1936
\cite{Lehmann1936}, it was modeled as an ideal mathematical object:
a rotational ellipsoid with a smooth surface and a homogeneous and
isotropic internal structure. But over the last decades the quality
and the amount of geophysical observations have increased so that
the IC image has become more complex and even contradictory
\cite{Souriau2007,Deguen2012}. In this paper we show that many of
the disputed issues and unresolved difficulties related to the IC
can be simultaneously simplified or even solved by the hypothesis
of an eastward displacement of the IC by tens of kilometers from
the Earth's center in the equatorial plane.

If the fluid in the OC is at rest, then the equilibrium position of
the IC at the Earth's center of mass is imposed by its gravitational
interaction with the rest of the Earth. A permanent shift from this
static equilibrium position is possible only if other forces act on
the IC. In Sect.~\ref{flow}, using the recent results of numerical
simulations of the geomagnetic dynamo \cite{Aubertetal2013}, we
tentatively identify the perturbing forces with those describing the
electromagnetic and hydrodynamic interactions of the IC with the
asymmetric convective flow in the OC. Then in Sect.~\ref{nonhydro}
we propose the global forcing of a decentered IC as a source of the
nonhydrostaticity of the Earth's shape. Finally, in
Sect.~\ref{slichter} we show that our hypothesis is supported by the
long standing observational difficulties in measuring the effects of
the translational oscillations of the IC about the static
equilibrium position (the Slichter mode). In conclusion, we
conjecture that the IC participates and interacts with the turbulent
convective flow in the OC being driven into a chaotic motion with
time scales similar to those of the secular variations of the
geomagnetic field near the equator.

In the next three sections we leave aside the dynamic considerations
and using recent seismic observations we obtain an estimation of the
present position of the IC. In Sect.~\ref{resid} we resume the
results presented in \cite{VamosandSuciu2011} where we analyzed the
isotropic hemispheric asymmetry at the top of the IC (ATIC) of the
travel times of the PKIKP seismic waves which propagate through the
layer of 100~km thick below the inner core boundary (ICB). Using ray
theory we numerically computed the PKIKP travel times for an
eccentric position of the IC (Appendix~\ref{appnumer}) and we showed
that the displacement of the IC toward 110$^{\circ}$~E in the equatorial
plane by tens of kilometers from the Earth's center can explain both the
values and the geographical distribution of the differential travel
times of the PKiKP-PKIKP phases. By means of the same numerical
model, in Sect.~\ref{atenua} we show that the decentering of the IC
also explains the hemispheric asymmetry of the attenuation within
the same region of the IC. So both hemispheric seismic asymmetries
at the top of the IC are simultaneously explained by geometrical
considerations related to the displacement of the IC, without
disturbing the symmetry of the internal structure of the IC.

The differential travel times used to analyze the ATIC are not
affected by the large perturbations due mainly to mantle
heterogeneity. In Sect.~\ref{precrit} we use the absolute travel
times of the seismic waves reflected at the ICB under small angles
and we document for the first time the likely existence of an
hemispheric asymmetry which could be related to a decentered
position of the IC. The result is not as rigorous as those obtained
in the previous two sections because we do not eliminate the
perturbations due to mantle heterogeneity, positioning errors of the
earthquake epicenters, or misidentification of the precritical PKiKP
phase.

An eccentric IC should influence many geophysical phenomena which
can be used as additional means to substantiate our hypothesis.
Section~\ref{geophys} contains qualitative analyses of some of these
geophysical phenomena: the anomalous layer above the ICB; the
differential rotation of the IC with respect to the mantle; the
large-scale anomalies of the geoid; the hemispherical pattern in the
anisotropy level of the seismic waves velocity inside the IC.

The hypothesis of an IC displaced by 100~km from the Earth's center
was formulated by Barta since the 1970's based on the distributions
of the large-scale anomalies of the geoid
\cite{barta1973,Barta1974,barta1981}. Other two more qualitative
justifications of this hypothesis were formulated in the same period
and are briefly presented in Appendix~\ref{apphist}. The decentering
of the IC has also recently been derived as a consequence of the
theory of translational convection inside the IC
\cite{Alboussiereetal2010,AlboussiereandDeguen2012}, but the
supposed displacement is of only 100~m, so that the IC remains in
mechanical equilibrium (see Appendix~\ref{apphist}).

\section{Dynamics of a decentered IC\label{interact}}

The main obstacle to accept the possibility that the IC could be
decentered by tens of kilometers is the identification of forces
large enough to displace the IC from its static equilibrium
position. In this section we explore some of the evidences
supporting the existence of such forces and of their dynamical
effects on the IC. The ability of the IC to be driven by the flow
inside the OC has to be tested through direct numerical simulations.

\subsection{Flow and geomagnetic field inside the OC\label{flow}}

Mechanical and electromagnetic interactions between the IC and OC,
together with the very presence of the IC, affect the geometry of
the flow inside the fluid OC and the geomagnetic field. The
decentering of the IC is an additional geometrical and mechanical
forcing which influences the structure of the magnetic field. In the
following we try to identify the features of the geomagnetic field
which support the existence of such an asymmetric forcing inside the
Earth. First we briefly present the basic structure of the
geomagnetic field which can be explained by symmetric numerical
geodynamo models.

The radial components of the geomagnetic field at the core-mantle
boundary (CMB) can be computed from the magnetic field at Earth's
surface assuming that no sources exist between the Earth's surface
and the CMB \cite{olsen2007}. The imaginary cylinder coaxial with
Earth's rotation axis and tangent to the IC at the equator, known as
the tangent cylinder, separates two regions of the OC in which the
fluid flow and the resulting magnetic field are quite different. The
most prominent features in the maps of the vertical field at the CMB
are the high intensity flux lobes under Arctic Canada, Siberia, and
under the eastern and western edges of Antarctica
\cite{jackson2007}. These lobes give the predominantly dipole field
structure observed at the surface which has remained approximately
stationary over the past four centuries. The flow near the top of OC
derived from the magnetic field at the CMB using a kinematic model
also contains polar vortices \cite{holme2007}.

There are many numerical dynamo models which can explain the main
properties of the geomagnetic field
\cite{christen2007}. For example, numerical simulations generate
weaker field inside the tangent cylinder, strong normal polarity
flux lobes close to the tangent cylinder, and pairwise inverse field
patches around the equator. An outstanding difficulty for standard
models has been how to reproduce the westward drift of low-latitude
magnetic flux patches at the CMB. At Earth's surface they correspond
to field variations with timescales shorter than 400 years, between
20$^{\circ}$~N and 20$^{\circ}$~S with
speed of approximately 17 km/yr westward \cite{jackson2007}.

A simple geometrical reasoning shows that the IC could be the cause
of this phenomenon. If the IC is displaced in the equatorial plane,
then its mechanical and thermal forcing on the OC is concentrated
between two planes parallel with the equator and tangent to the IC,
i.e., within a spherical segment with the height equal with the IC
diameter of 2442~km. This height differs only by a few percents from
the width at the CMB of the westward-moving wave-like patterns equal
to $2R_{\mathrm{OC}}\sin 20^{\circ}=2380\,\mathrm{km}$, where
$R_{\mathrm{OC}}=3480\,\mathrm{km}$ is the OC radius.

A more elaborate justification is provided by the first successful
numerical simulation of the westward drift of the magnetic field,
obtained by means of a heterogeneous thermodynamical boundary
condition at both the ICB and CMB and a gravitational coupling of
the IC with the mantle \cite{Aubertetal2013}. This dynamo model
produces magnetic variations dominated by intense,
westward-drifting, equatorial flux patches under the Atlantic
hemisphere. The maximum of the magnetic power moving in the
longitudinal direction is reached at the Equator, with a maximum
speed of 14~km/yr, comparable to that observed of 17~km/yr.

The numerical simulated convection in the OC is dominated by the
mass flux at the ICB modeled by a longitudinal hemispherical
heterogeneity, maximal at the longitude of $90\,^{\circ}\mathrm{E}$.
This asymmetric forcing could be caused by the displacement of the
IC in the direction of the maximum mass flux. Because of the smaller
distance between the ICB and the CMB, in this direction the
temperature gradient should be larger and the mass flux should be
increased. The gravitational coupling between the IC and the mantle
could be a secondary effect of the displacement of the IC.

An important result obtained by this numerical simulation related to
the IC dynamics is the strong large-scale asymmetry of the flow and
of the magnetic field inside the OC. The rotational turbulent
hydrodynamic flow in the OC can become asymmetric at planetary
scale, firstly by interaction with the IC and secondly by the
heterogeneous forcing at the CMB. The asymmetry of the flow and of
the magnetic field at the CMB is also supported by the observation
data \cite{Galletetal2009,Hulotetal2002}. Such a complex flow could
generate significant viscous and magnetic forces on the IC
\cite{Aubertetal2008,BuffettandBloxham2000,BuffettandGlatzmaier2000}
and could maintain it in a permanent chaotic motion. These forces
should increase when the nondimensional parameters characterizing
the numerical simulations will further approach the real values for
the deep Earth. The time scales characteristic to the IC motion have
to be larger than several decades which is the time scale associated
to the westward drift of the geomagnetic field.

The relation between the direction of the highest light-element mass
flux at the ICB and the offset of the magnetic dipole from the
Earth's center \cite{OlsonandDeguen2012} also supports our
hypothesis of a decentered IC. The drastic change in the dipole
position from the Western to the Eastern Hemisphere during the last
10000 years could be explained by a complete reversal of the
direction of the faster IC growth, which in turn could result from
intermittent IC rotation \cite{OlsonandDeguen2012}. A simpler
explanation might be the movement of the IC from the Western to
Eastern Hemisphere, which reverses the direction of the maximum mass
flux associated with the hemispheric asymmetry of the magnetic field
\cite{Aubertetal2013}. This provides an upper limit of several
thousands of years for the time scales of the IC motion. In
addition, the intricate trajectory of the dipole is an indication
that the motion of the IC is likely chaotic.

\subsection{The variation of the Earth's rotation\label{nonhydro}}

The dynamic forcing of a decentered IC should be transmitted to the entire
Earth. Here we briefly discuss its influence on the rotational
motion of the Earth and on the nonhydrostaticity of the Earth's
shape. Variations of the Earth rotation manifest as variations in
direction of Earth's rotation axis as well as variations in the angular
speed, i.e., variations in the length of day (LOD). Precession of an
axis is the mean, smoothly varying part of the motion of the axis
relative to the direction of fixed stars while nutation is the
oscillatory part of this motion \cite{dehant2007}. The motion of the
instantaneous rotation axis with respect to the figure of the Earth
is known as the Earth wobble.

The influence of the IC on the wobble and nutation of the Earth can
be determined from a three-layer model of a deformable Earth
\cite{dehant2007}. The results of the numerical model are a set of
optimized values for the Earth parameters derived from the ``best
fit'' with observational data. We are first interested in the
inference that the ellipticity of the core has to be about 5\%
higher than its value for the hydrostatic equilibrium Earth model
\cite{gwin86}. The nonhydrostatic excess of the ellipticity
corresponds to a difference of approximately 400~m between the
equatorial and polar radii of the CMB.

This nonhydrostaticity has been attributed to the mantle convection
\cite{dehant2007}. Another cause could be that the IC displacement
perturbs not only the OC, but also the mantle, mainly in the
equatorial region. Our estimations indicate a displacement of the IC
from the Earth's center in the equatorial plane of tens of
kilometers (Sects.~\ref{resid}, ~\ref{atenua}, and ~\ref{precrit}).
This is almost two orders of magnitude larger than the displacement
deduced from mechanical and thermal equilibrium conditions
\cite{Alboussiereetal2010}. Such a large difference implies a
nonhydrostatic evolution of the IC and, implicitly, of the rest of
the Earth.

Another phenomenon possibly influenced by the IC displacement is the
Makowitz wobble, which is the motion of the pole with respect to the
Earth's crust and mantle including a quasiperiodic component with a
period of approximately 24 years superimposed on a linear drift
\cite{gross2007}. The cause of the decadal-scale polar motion
variations is currently unknown. It was found that the main
excitation source of the variations cannot be the redistribution of
mass within the atmosphere and oceans or the core-mantle coupling.
The most probable source is the IC, but the suggestion that
irregular motion of a tilted oblate IC may excite the Makowitz
wobble was proved to be unlikely. An eccentric IC could have the
additional independent parameters needed to explain this wobble.

The rotation period of the Earth is not uniform, but varies on time
scales from days to millennia \cite{gross2007}. The tidal drag of
the Moon and Sun on the rotating Earth produces a secular slowing
down of the rotation. The angular momentum variations of the
atmospheric and oceanic global circulation explain most of the
observed LOD variation at yearly and subyearly timescales. In
addition, there are variations over several decades related to the
angular momentum in the core which are well modeled for the past
century \cite{jackson1993}. Prior to this period the results are
poorer, especially because of a phase shift, but the general pattern
is still remarkable. The supposed displacement of the IC would
affect the global angular momentum of the Earth and could be an
explanation of the phase shift. Then the dominant time scales of the
IC movement could be at least several hundreds of years.

\subsection{Translational oscillations of the IC\label{slichter}}

An equilibrium position of the IC at the Earth's center of mass is primarily determined by the
gravitational field. The gravitational restoring force is
proportional with the displacement of the IC from the Earth's center
and with the density difference between the IC and the surrounding
OC. A large earthquake or a large meteorite impact could
initiate harmonic oscillations about this equilibrium position. If
$\omega_0$ is the angular frequency for oscillations parallel to the
Earth's rotation axis, then the oscillations in the equatorial plane
have two eigenvalues $\omega_0 \pm \Omega$, where $\Omega$ is the
Earth's angular velocity. The three modes of translational
oscillations are known as Slichter modes after the researcher who
first predicted them \cite{slichter61}.

The frequencies of the Slichter modes are influenced by many
factors: the stratification of the fluid in the OC, the elastic
properties of the IC and OC, the viscosity of the fluid in the OC,
the magnetic perturbations near the ICB, the mushy structure of the
ICB, etc.~\cite{rosat2004}. Depending on the approximations used to
model these factors, the theoretically computed periods vary from 4
to 8~h according to different authors. But this mode has never been
clearly identified from observational data and it is still a subject
of interest and debate. The Slichter mode was not observed even in
the records obtained after the huge Sumatra-Andaman earthquake of
December 2004 \cite{Okal2009,roult2010}.

It is obvious that the inability to detect the Slichter mode could
be due to the insufficient sensitivity of the currently available
superconducting gravimeters. However, taking into account that all
the other normal modes of the Earth have been identified from the
seismic data, it is possible that the translational oscillations
simply do not take place because the gravitational restoring force
is not the only force acting on the IC. If, as we have  conjectured
in Sect.~\ref{flow}, the IC is in a chaotic motion correlated to the
turbulent flow in OC, then the momentum transferred by an earthquake
would not cause the oscillation of the IC, but it would determine
only an additional displacement. Hence the difficulty to determine
the Slichter mode could be an indication that the IC is not in an
equilibrium position at the Earth's center.

\section{Residuals of the PK\lowercase{i}KP-PKIKP differential travel time\label{resid}}

In this section we begin to determine the present position of the IC
using seismic data. One of the best documented seismic hemispheric
asymmetry is the isotropic asymmetry at the top of the inner core
(ATIC) characterized by an Eastern Hemisphere with faster arrival
times of the P-waves and a Western Hemisphere with slower arrivals
\cite{GarciaandSouriau2000,OuzounisandCreager2001,
NiuandWen2001,WenandNiu2002,CaoandRomanowicz2004,Yuetal2005,YuandWen2006}.
ATIC has been mainly documented by the residuals of the differential
travel time of the PKIKP and PKiKP seismic phases
\cite{NiuandWen2001,WenandNiu2002,Yuetal2005,YuandWen2006,Waszeketal2011,WaszekandDeuss2011}.
They both travel through almost the same regions of the crust,
mantle, and OC. After that, the PKiKP phase reflects off the ICB,
while the PKIKP phase refracts twice on ICB propagating inside the
IC.

We denote by $\Delta t$ the observed differential travel time
obtained by subtracting the travel time of the PKIKP phase from the
travel time of the PKiKP phase with the same focus and exit point.
It differs from the differential travel time $\Delta t_0$ computed
for a centered IC with the velocity profile given by a
one-dimensional reference seismic model. The observational data show
that the residuals $\Delta t-\Delta t_0$ are positive in the Eastern
Hemisphere and negative in the Western Hemisphere
\cite{NiuandWen2001,WenandNiu2002,Yuetal2005,YuandWen2006,Waszeketal2011}.
An important constraint for the models proposed to explain this
hemispheric dichotomy is the sharpness of the boundaries separating
the regions with positive and negative residuals
\cite{Waszeketal2011,WaszekandDeuss2011}.

All existing explanations assume the center of the IC fixed at the
center of the Earth and interpret the observed anomaly of the travel
time residuals in terms of a longitudinal anomaly of the seismic
wave velocity (e.g.~\cite{WaszekandDeuss2011}). Greater (smaller)
seismic wave velocity at the top of the IC in the Eastern (Western)
Hemisphere, with respect to 1D reference models, are explained by a
hemispheric variation of the material properties at the top of the
IC
\cite{NiuandWen2001,WenandNiu2002,Yuetal2005,YuandWen2006,Alboussiereetal2010,Waszeketal2011,WaszekandDeuss2011,Cormieretal2011}.
There are two competing approaches to explain such an IC velocity
asymmetry. The first one assumes different cooling rates in the
Eastern and Western Hemispheres due to thermochemical coupling with
the mantle
\cite{SumitaandOlson1999,SumitaandOlson2002,Aubertetal2008}, causing
a faster solidification rate of the Eastern Hemisphere. Different
textures of the IC material resulted from this process may explain
the hemispherical pattern of the seismic velocity
\cite{Cormier2007}. Alternatively, another approach proposes a
self-sustained eastward translation of the IC as a result of
crystallization in the Western Hemisphere and melting in the Eastern
Hemisphere, followed by the west-east increase of the iron
grain-size which could produce the velocity anomaly explaining the
ATIC \cite{Alboussiereetal2010,Monnereauetal2010}. Even whether
these models explain the travel time anomaly, neither of them is
fully consistent with the observed sharpness of the hemispheric
boundaries \cite{WaszekandDeuss2011,AlboussiereandDeguen2012}.

\begin{figure}
\includegraphics{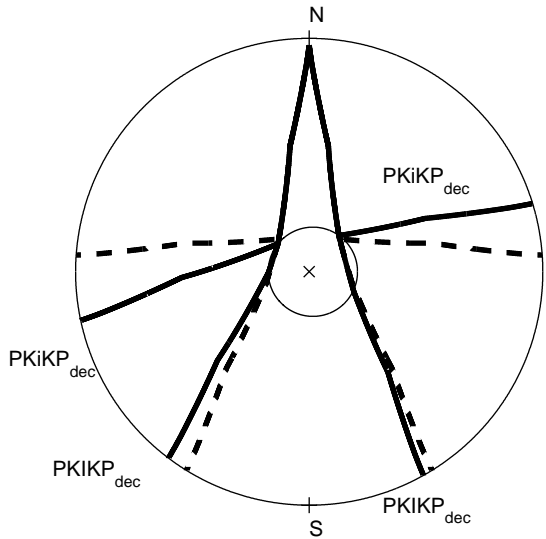}
\caption{\label{asim}The paths of the PKIKP$_{\mathrm{dec}}$ and
PKiKP$_{\mathrm{dec}}$ rays (thick continuous lines) when the IC
(the small circle) is displaced by 100~km from the Earth's center
in the equatorial plane toward east. The epicenter is at the North
Pole, the propagation plane of the seismic rays contains the centers
of both Earth and decentered IC and, in this case, they are not
modified by refraction or reflection on ICB. The dashed lines
represent the paths of the same phases for the centered IC.}
\end{figure}

Using the numerical simulation presented in Appendix~\ref{appnumer}
we show that the ATIC can be explained by the displacement of the IC
in the equatorial plane toward east by tens of kilometers from the
Earth's center, without modifying the spherical symmetry in the
upper IC. We denote by PKIKP$_{\mathrm{dec}}$ and
PKiKP$_{\mathrm{dec}}$ the paths modified by the decentered IC.
Unlike the paths for the centered IC, their propagation plane
changes at reflection or refraction on ICB. Only when the seismic
ray propagates in a plane containing both the center of the IC and
of the Earth, the propagation plane does not change. A clear
east-west asymmetry is obvious in Fig.~\ref{asim} for such seismic rays
having the same initial incidence angle, i.e., being identical until
the incidence with the ICB.

\begin{figure}
\includegraphics{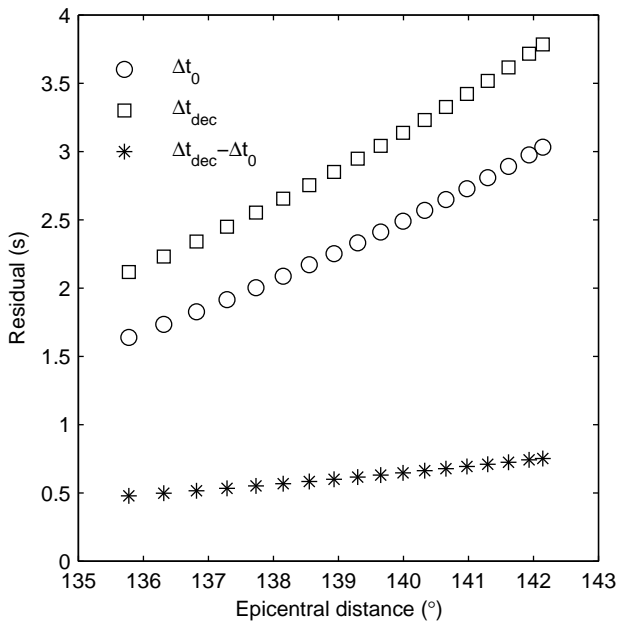}
\caption{\label{difft}Numerically computed residuals for a
decentered IC. The differential travel times $\Delta
t_{\mathrm{dec}}$ are computed for the PKIKP$_{\mathrm{dec}}$
seismic rays in the Eastern Hemisphere plotted in Fig.~\ref{asim}.}
\end{figure}

From the differential times computed in Appendix~\ref{appnumer} we
determine the residuals shown in Fig.~\ref{difft} for the seismic
rays in the Eastern Hemisphere plotted in Fig.~\ref{asim}. They are
quantitatively comparable with those observed
\cite{NiuandWen2001,WenandNiu2002,Yuetal2005,YuandWen2006,Waszeketal2011}
showing that displacements of the IC over distances up to 100~km can
explain the travel time anomaly. With the increase of the turning
point depth (epicentral distance) the positive residuals in the
Eastern Hemisphere increase because the length of the path inside
the IC increases (Fig.~\ref{icbray}a).

In order to ascertain if a decentered IC can explain ATIC, we
compare the longitudinal repartition of the residuals obtained by
numerical simulations with those reported in \cite[Fig.~3a]{Waszeketal2011}
and in \cite[Fig.~3b]{WaszekandDeuss2011}, the most
extensive and accurate PKiKP-PKIKP studies to date. For a given
position of the IC, we compute the residuals for earthquakes evenly
distributed on the Earth's surface and seismic rays uniformly
distributed around the focus (see Appendix~\ref{appnumer}). For each
seismic ray we choose the incidence angle so that the corresponding
turning point is at 39~km below ICB, the minimum depth below ICB of
the available observational data (see \cite[Fig.~3a]{Waszeketal2011}).

\begin{figure*}
\includegraphics{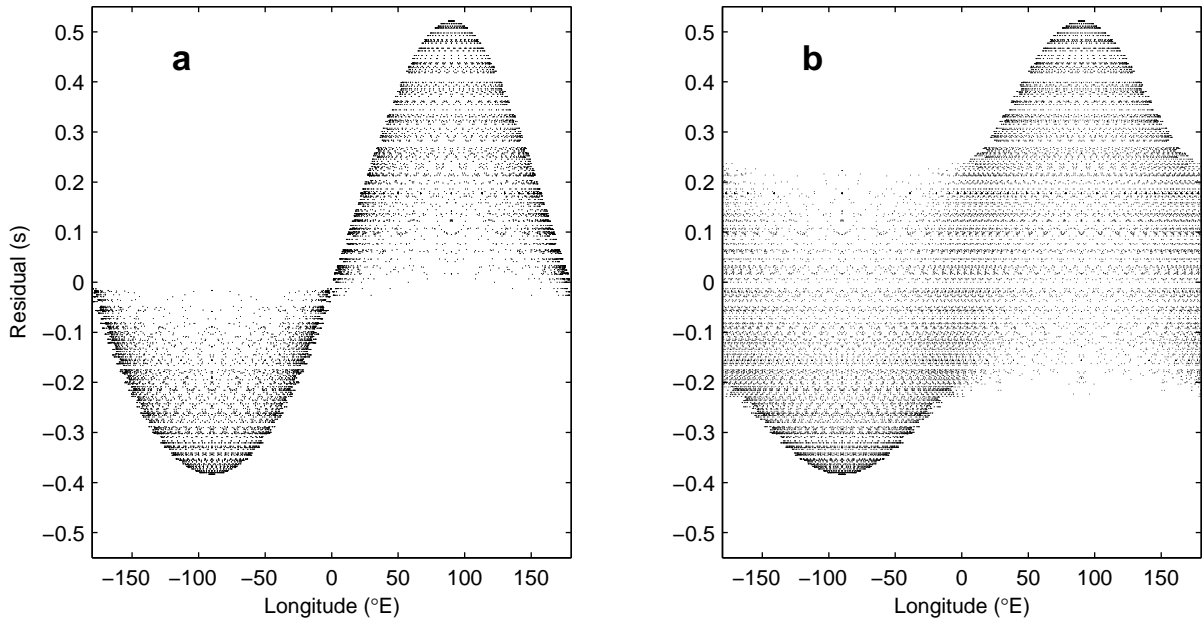}
\caption{\label{reprez}Longitudinal distribution of the ATIC
residuals obtained by numerical simulation. The IC is displaced by
100~km toward 90$^{\circ}$~E longitude in the equatorial plane (a)
and along a direction making an angle of 30$^{\circ}$ with the
equatorial plane (b). The abscissa represents the longitude of the
turning point of the PKIKP$_{\mathrm{dec}}$ ray.}
\end{figure*}

Figure~\ref{reprez} shows the longitudinal distribution of the
residuals when the IC is displaced by 100 km toward 90$^{\circ}$~E
longitude. If the displacement is in the equatorial plane, then the
positive residuals are confined within the Eastern Hemisphere and
the negative ones within the Western Hemisphere
(Fig.~\ref{reprez}a). If the IC is displaced outside the equatorial
plane, the separation of the positive and negative residuals is not
so definite (Fig.~\ref{reprez}b). In the observational data, the
positive and negative residuals are sharply separated
\cite{Waszeketal2011,WaszekandDeuss2011} indicating that the
displacement of the IC is in the equatorial plane
(Fig.~\ref{reprez}a). The observed boundary between the positive and
negative residuals is shifted toward east by approximately
20$^{\circ}$ with respect to the boundary between the Eastern and
Western Hemispheres. The numerically simulated boundary rotates with
the angle equal with the difference between the longitude of the IC
center and the 90$^{\circ}$~E longitude. All these
indicate a displacement of the IC by tens of kilometers in
equatorial plane toward 110$^{\circ}$~E longitude.

The theoretical residuals generated by a decentred IC have a
cylindrical symmetry about the direction of the IC displacement.
This distribution is consistent with the eyeball-shaped positive
anomaly at low- and mid-latitude of the compressional velocity
derived for a centered IC model  from seismical observations
\cite{tanaka1997}. The anomaly is centered on 90$^{\circ}$~E
longitude on the equator, i.e., very close to the direction of the
IC displacement estimated above. The theoretical distribution of the
residuals also agrees with the lack of the hemispheric asymmetry at
the south pole of the IC \cite{ohtaki2012}.

When the results of the numerical simulations are compared with
observational data we have to take into account the simplifying
hypotheses of the numerical simulation as well as the observational
errors. For instance, we have used the velocities of the ak135 model
obtained under the hypothesis that the IC is centered. But the
observational differential travel times PKiKP-PKIKP are spread
around the mean values of the ak135 model by about 0.5~s
\cite{Kennettetal1995}, which is comparable with the values
associated with ATIC \cite{Waszeketal2011}. That is why the exact
longitude separating the positive and negative observed residuals
and its variation with the turning point depth of the seismic rays
cannot be determined precisely. The observation data suggest an
eastward shift of the hemisphere boundary with increasing depth
\cite{Waszeketal2011}, while our numerical simulations have shown
that it does not vary with the depth.

\section{Hemispheric asymmetry of the attenuation in the upper IC\label{atenua}}

There are other seismic phenomena with east-west asymmetry, although
without a complete observational description of their longitudinal
variation, which may be explained by a decentered IC. For instance,
ATIC is associated to a hemispheric asymmetry of the seismic waves
attenuation
\cite{WenandNiu2002,LiandCormier2002,CaoandRomanowicz2004,
OreshinandVinnik2004,YuandWen2006,Souriau2007,Iritanietal2010} which
seems to be confined to the uppermost IC
\cite{LiandCormier2002,Souriau2007}. Existing explanations of the
attenuation asymmetry require a trade-off between attenuation and
velocity structures in the IC and velocity structure at the bottom
of OC \cite{WenandNiu2002,YuandWen2006}. Explanations of the
hemispheric asymmetric attenuation are currently based on the
specific texture of the uppermost IC \cite{Cormier2007} or on the
assumption that a mushy zone exists at ICB
\cite{CaoandRomanowicz2004}.

If the IC is decentered, the PKIKP$_{\mathrm{dec}}$ phase propagates
in the Eastern Hemisphere over a longer distance inside the IC
(segment CD in Fig.~\ref{icbray}a) than in the Western Hemisphere.
Since the quality factor $Q$ is two orders of magnitude larger in
the OC than in the IC \cite{Kennettetal1995}, the attenuation
$Q^{-1}$ in the Eastern Hemisphere is larger than in the Western
Hemisphere. Hence the different lengths of the
PKIKP$_{\mathrm{dec}}$ paths in Eastern and Western Hemispheres of
the decentered IC explain not only the hemispherical asymmetry
travel time residuals (Sect.~\ref{resid}), but also the hemispheric
asymmetry of the attenuation.

The quality factor $Q$ at the top of the IC is determined by
\[
\frac{A_{\mathrm{I}}}{A_{\mathrm{i}}}=exp\left\{ -\frac{\pi
ft}{Q}\right\},
\]
where $A_{\mathrm{I}}$ and $A_{\mathrm{i}}$ are the amplitudes of
the PKIKP and PKiKP phases at the common emerging point, $f$ is the
frequency and $t$ is the travel time of PKIKP phase inside the IC
\cite{CaoandRomanowicz2004}. The quantities $A_{\mathrm{I}}$,
$A_{\mathrm{i}}$, and $t$ are derived from measurements and the
above relation is used to compute the quality factor $Q$.

The seismical observations show that the ratio
$A_{\mathrm{I}}/A_{\mathrm{i}}$ depends not only on the epicentral
distance $\Delta$, but also on the geographic location. If one
considers that the IC is centered, then the travel time inside the
IC depends only on the epicentral distance $t_{\mathrm{cen}}(\Delta
)$. In this case, the geographical variations of
$A_{\mathrm{I}}/A_{\mathrm{i}}$ are possible only if the material
properties vary inside the IC and the quality factor depends, besides
the epicentral distance, also on the longitude and
latitude of the turning point H, $Q_{\mathrm{cen}}(\Delta
;\lambda_H,\phi_H)$. The quality factor $Q_{\mathrm{cen}}$ computed
for a centered IC
takes larger values in the Western Hemisphere than in the Eastern
Hemisphere, the difference being larger at smaller epicentral
distances \cite[Fig.~5b]{CaoandRomanowicz2004}.

In the following, we show how a displacement over 100~km of the IC in
the equatorial plane can explain the seismical observations, without
the need to modify the quality factor inside the IC.
Using the numerical model presented in Appendix~\ref{appnumer}, we
compute the travel time of the PKIKP phase in a decentered IC,
$t_{\mathrm{dec}}(\Delta ;\lambda_H,\phi_H)$, which is a function of
the longitude and latitude of the turning point H. We describe the
attenuation at the top of the IC by means of a constant
quality factor $\overline{Q}=247$ \cite{CaoandRomanowicz2004}. Then
the two models of attenuation are related by the relation
\[
\frac{t_{\mathrm{cen}}(\Delta )}{Q_{\mathrm{cen}}(\Delta
;\lambda_H,\phi_H)}=\frac{t_{\mathrm{dec}}(\Delta
;\lambda_H,\phi_H)}{\overline{Q}} .
\]
If our hypothesis is correct, then the quantity $Q_{\mathrm{cen}}$
computed from this formula should be identical with that derived
directly from observational data.

\begin{figure}
\includegraphics{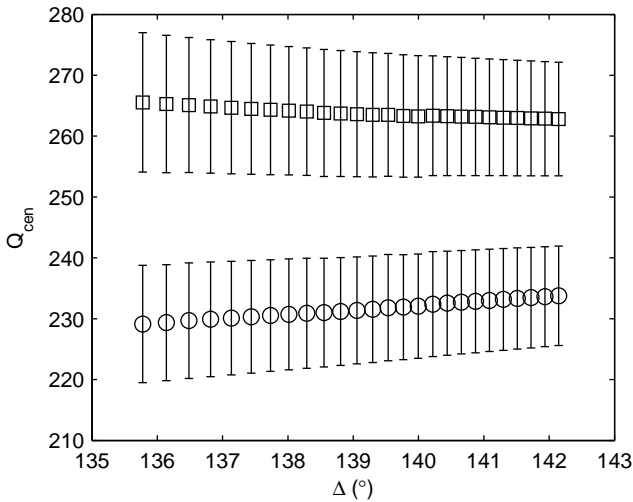}
\caption{\label{aten}Attenuation for a decentered IC. Average and
standard deviation over the Eastern (circle markers) and Western
(square markers) Hemispheres of the simulated quality factor
$Q_{\mathrm{cen}}$ as a function of the epicentral distance.}
\end{figure}

We have computed the mean and the standard deviation of
$Q_{\mathrm{cen}}$ over all possible geographical positions (angles
$\lambda_H$ and $\phi_H$), separately for east longitudes
($\lambda_H>0$) and west longitudes ($\lambda_H<0$). The numerical
results presented in Fig.~\ref{aten} show a good resemblance with
the observational results (see
\cite[Fig.~5b]{CaoandRomanowicz2004}). The distance between the two
branches of the graph decreases when the epicentral distance
increases in the same way as for the observed data, although they do
not intersect at epicentral distances greater than 140$^{\circ}$.
Also the observed differences for epicentral distances smaller than
140$^{\circ}$ are several times greater than those numerically
computed.

\section{Hemispheric asymmetry of the precritical PK\lowercase{i}KP travel time\label{precrit}}

A displacement of the IC should give rise to other
hemispheric asymmetries of the seismic phases propagating through
the IC. Unfortunately, due to the inhomogeneities of the Earth's
interior, the absolute travel times for such seismic rays reaching
the ICB are difficult to be determined and then the geographical
repartition of the characteristics of the seismic waves is highly uncertain.
The largest changes of the arrival times induced by the position of
the ICB occur for the seismic rays reflected under small angles
(near-normal PKiKP phase). Identification of such perturbations in
seismic observations encounters two major difficulties.

First, in accordance with recent studies, the surface of the ICB has
a rough topography \cite{Koperetal2003} with height variations
larger than 10 km which produce local fluctuations of several
seconds of the near-normal PKiKP travel times \cite{Daietal2012}.
Thus a displacement of the entire IC over a distance of the same
order of magnitude is hidden by the local fluctuations and a large
enough sample of data is needed to assure a significant statistical
analysis. The second problem is related to the difficulty to
identify the near-normal PKiKP phase from seismic observations
\cite{ShearerandMasters1990,Koperetal2003}. If the IC is decentered,
then the identification is even harder because the arrival times can
become quite different from those computed by a reference Earth
model with spherical symmetry, which are used in phase
identification procedures
\cite{Engdahletal1998,ShearerandMasters1990,Pengetal2008}. For
instance, a 10 km displacement of the IC, which is comparable to the
height of the ICB irregularities, causes a variation up to 2~s of
the arrival times of the near-normal PKiKP waves. Therefore we
expect that the near-normal PKiKP rays with the largest residuals
are most probably misidentified.

Rather than relying on particular events with well identified and
analyzed PKiKP phases, as for instance in
\cite{Koperetal2003,PoupinetandKennett2004,Kawakatsu2006,JiangandZhao2012},
we consider global data and follow a statistical approach. We use
validated ISC data \cite{ISC2013} to assemble a large enough sample
which allows us to draw significant statistical conclusions. We
downloaded information for all phases identified by ISC as PKiKP,
irrespective of how they were initially reported. While before 2006
the reported PKiKP phases were generally not confirmed by the ISC
analysts, since 2006 they identified the PKiKP phase and computed
arrival times residuals with respect to the ak135 velocity model (D.
A. Storchak, private communication, 2013). For our purpose, we
select data with $\Delta<90^{\circ}$, which cover the whole range of
precritical PKiKP and most of the transparent zone
\cite{Koperetal2003,Kawakatsu2006}.

\begin{figure*}
\noindent\includegraphics{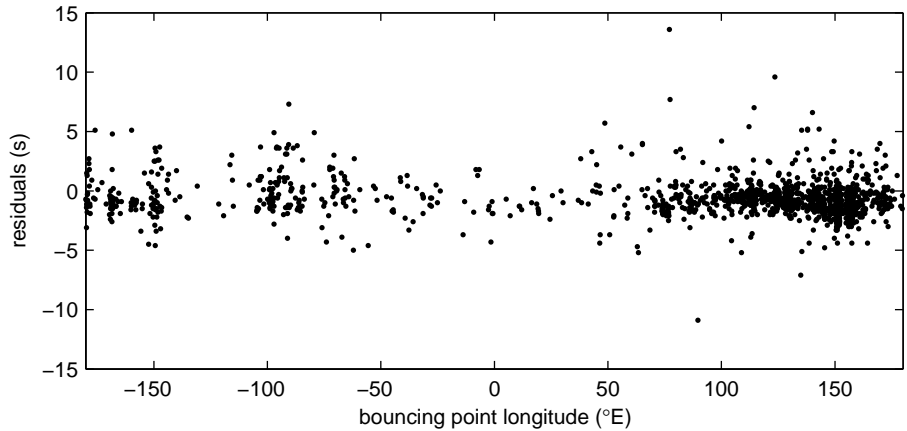} \caption{\label{precr}Arrival
time residuals  with respect to the reference model ak135 of the
PKiKP phase, recorded at epicentral distances smaller than
$90^{\circ}$, as a function of the longitude of the bouncing points
at the ICB.}
\end{figure*}

From the validated data set 2006-2010 we have obtained a sample of
2042 residuals of the precritical PKiKP phase, plotted in
Fig.~\ref{precr} as function of the longitude of the bouncing points
on the ICB. The residuals show large fluctuations similar to those
reported in other recent studies on PKiKP phase
\cite{Koperetal2003,Daietal2012}. They have a negative mean of
-0.745 s, with a standard deviation of 1.446 s and a standard error
of the mean of 0.032 s (estimated by standard deviation divided by
the square root of number of observations), in agreement with
previous results obtained from smaller data sets
\cite{Koperetal2003}. Removal of the outliers from the data sample
has insignificant influence on the statistical results derived in the
following.

The sample of 2042 residuals contains 1669 observations with the
bouncing point in the Eastern Hemisphere and only 373 observations
in the Western Hemisphere. In the Eastern Hemisphere the mean of the
residuals is -0.838 s, with standard deviation of 1.361 s and
standard error of the mean of 0.033 s, while in the Western
Hemisphere the mean residual is -0.328 s, with standard deviation of
1.720 s and standard error of the mean of 0.089 s. Thus, the data
set shows a difference of 0.510 s between the mean precritical PKiKP
residuals in the Western and Eastern Hemispheres, with a standard
error of 0.095 s (computed considering that the two subsets of
seismic events are independent). The west-est discrepancy of the
mean residuals, with faster precritical PKiKP arrivals in the
Eastern Hemisphere, indicates a global asymmetry not yet reported in
the literature.

\begin{figure}
\noindent\includegraphics{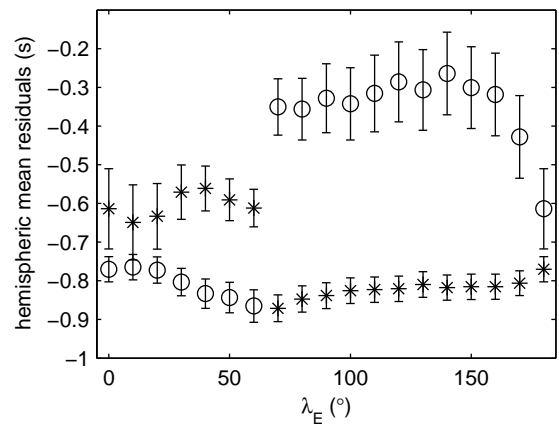} \caption{\label{emisf}Residuals
averaged over the $H_E$ hemisphere (asterisks) and over the $H_W$
hemisphere (circles) as a function of longitude $\lambda_E$ of the
middle of $H_E$. Error bars represent standard errors of the mean.}
\end{figure}

We assume that the maximum hemispheric asymmetry lies in the
equatorial plane, as we already found in the previous sections. We
search for a partition of the Earth into two disjoint hemispheres
which yields the maximum difference of the hemispheric averages of
the precritical PKiKP residuals. We denote by $\lambda_E \in
[0^{\circ},180^{\circ}]$ the eastern longitude of the middle of the
hemisphere $H_E$ containing the greater part of the geographic
Eastern Hemisphere. (We adopt the usual convention that eastern
longitudes are positive, while western ones are negative.) Hence,
the hemisphere $H_E$ is defined by the longitudes $\lambda \in
[\lambda_E-90^{\circ},\lambda_E+90^{\circ}]$ if $\lambda_E \leq
90^{\circ}$ and $\lambda \in [\lambda_E-90^{\circ},180^{\circ}] \cup
[-180^{\circ},\lambda_E-270^{\circ}]$ if $\lambda_E \geq
90^{\circ}$. We denote by $H_W$ the complement of $H_E$ containing
mostly western longitudes centered on the longitude $\lambda_W =
\lambda_E -180^{\circ}<0$. Figure~\ref{emisf} shows the mean
residuals and the standard errors of the mean for PKiKP rays with
bouncing points in the two hemispheres as a function of $\lambda_E$.
For $70^{\circ}\leq \lambda _{E}\leq 170^{\circ}$ the mean residuals
in the $H_E$ hemisphere are smaller than in the $H_W$ hemisphere by
a quantity several times larger than the standard errors of the
mean. The largest difference, equal to 0.55~s, is obtained for
$\lambda _{E}=140^{\circ}$. For a wave velocity of about 10~km/s,
this difference corresponds to a mean difference of 5.5~km between
the paths of the PKiKP rays in the two hemispheres. Hence, the $H_E$
hemisphere of the ICB  centered on $\lambda_{E}=140^{\circ}$ is on
average closer by 2.7~km to the Earth's surface than the
corresponding $H_W$ hemisphere.

In Fig.~\ref{emisf} our attention is drawn by the sudden reversal of
the sign of the mean residuals when passing from $\lambda
_{E}=60^{\circ}$ to $\lambda _{E}=70^{\circ}$. The jump occurs when
the plane separating the two hemispheres rotates by 10$^{\circ}$ and
the residuals with bouncing point longitude between 150$^{\circ}$E
and 160$^{\circ}$E move from the $H_W$ hemisphere to the $H_E$
hemisphere. There are 589 such observations, i.e., 29\% from the
total number of observations (see Fig.~\ref{precr}). The sign
reversal shows that these data are responsible for two thirds of the
maximum difference of the mean residual obtained for
$\lambda_E=140^{\circ}$. The remaining one third is due to the other
residuals in the $H_E$ hemisphere.

In order to confirm the existence of this new hemispherical seismic
asymmetry, additional work is needed. The influence of the site
effects, the near source heterogeneity, and the long wavelength
mantle heterogeneities should be computed and subtracted for each
travel time. Also the PKiKP phases reported by ISC should be
individually verified and validated. However, all these perturbing
effects are likely uncorrelated with the seismic events and they
might have been significantly reduced by the statistical processing of the
2042 observations. If confirmed, this asymmetry is an indication that the
value of 2.7~km should be an inferior limit for the IC displacement.

\section{Other geophysical implications of a decentered IC\label{geophys}}

There are many geophysical phenomena containing qualitative and
quantitative information regarding the position and the motion of
the IC which could support the hypothesis of a decentered IC. For
example, the bifurcation of the PKP phases occurring at smaller
epicentral distance in the Eastern than in the Western Hemisphere
\cite{Yuetal2005,YuandWen2006}. Also, the observation that travel
time residuals of the PKiKP with respect to the
PKPB$_{\textrm{diff}}$ phase, diffracted through the middle of the
OC, are smaller by about 0.9~s in the Eastern Hemisphere than in the
Western Hemisphere \cite{Yuetal2005} is consistent with the PKiKP
travel time hemispheric asymmetry presented in Sect.~\ref{precrit}
and can be explained as well by the eccentric position of the IC.

Many of these geophysical phenomena with different types of
asymmetries are related to the structure and evolution of the
geomagnetic field. Among them the hemispheric asymmetry of the
magnetic field \cite{Hulotetal2002} which is responsible
for a more intense radial field at the longitude of about 150$^\circ$~E
\cite{Galletetal2009}, the same as the current longitude
of the eccentric dipole \cite{OlsonandDeguen2012} and close to that of
the hypothesized IC displacement. Movements of the IC in the equatorial
plane could also be related to the gravity variations induced by core
flows \cite{Dumberry2010grav} and the correlations between magnetic
and gravity anomalies at low and middle latitudes recently reported by
Mandea et al. \cite{Mandeaetal2012}.

In the following we discuss in more detail four such geophysical
phenomena and we show that the hypothesis of a decentered IC could
contribute to their explanation.

\subsection{The anomalous layer at the bottom of the OC\label{anomal}}

At the bottom of the OC seismological data reveal an anomaly
consisting of PKPbc-PKIKP residuals smaller than predicted by
reference one-dimensional models \cite{Yuetal2005}. This anomaly has
been interpreted as a layer of low velocity gradient in the
lowermost 150-200 km of the OC \cite{Souriau2007,Yuetal2005}, which
could delay the PKIKP travel time and produce smaller PKPbc-PKIKP
residuals \cite{Yuetal2005}. The favored explanation of this anomaly
is the existence of a stably stratified region resulting from a
dynamic equilibrium between the production of iron reach melt and
mixing with light fluid \cite{Deguen2012,Souriau2007}. Melting may
occur through the formation of a positive topography at the ICB by
translational IC convection
\cite{Alboussiereetal2010,Monnereauetal2010}, provided that the OC
convection supplies the necessary latent heat, or by OC convection
itself, when the temperature in the surrounding OC fluid is locally
larger than at the ICB \cite{Gubbinsetal2011}. However, both
mechanisms suffer of limitations and there is no definitive proof
that they are self-sustaining \cite{Deguen2012}.

The hypothesis of an eccentric IC could contribute to more
appropriate explanation of this anomalous layer from a new
perspective. The IC displacement over approximately 100~km should
produce noticeable effects at the bottom of the OC. If $r_c$ is the
radius of the IC, then the distances from the Earth's center to the
ICB vary between $r_c-100$ and $r_c+100$. Hence the observed
anomalous layer contains this spherical shell with an inhomogeneous
material structure which should be taken into account to explain the
seismical data. For example, the observed PKPbc-PKIKP and
PKiKP-PKIKP differential travel time anomalies, which can be
simultaneously explained by different radial velocity gradients at
the bottom of the OC in the two hemispheres
\cite{Yuetal2005,YuandWen2006}, could be related to the different
lengths of the PKIKP path inside a decentered IC.

There may be other seismic observations related to this region of
the OC. For instance, if the IC is decentered, while the seismic
model assumes that it is centered, then the interpretation of the
seismic observations would have larger errors in the spherical layer
of 200~km containing the ICB than in other regions of the OC.
Indeed, the reference one-dimensional seismic models
are different from each other over a thickness of roughly 200~km
above ICB \cite{Yuetal2005,YuandWen2006,Souriau2007}.

\subsection{Differential rotation of the IC and the hemispheric seismic asymmetry}\label{diffrot}

The flow in the fluid OC and the angular momentum conservation
induce changes in the angular velocity and rotation axis of the IC
\cite{Aubertetal2008,Suetal1996,DumberryandMound2010}. The
differential rotation of the IC, first predicted from theoretical
considerations \cite{Gubbins1981}, was obtained by geodynamo
numerical simulations in the mid-nineties
\cite{GlatzmaierandRoberts1996}. But the differential rotation of
the IC is hardly compatible with the current explanations of the
hemispheric dichotomy (see Sects.~\ref{resid} and \ref{atenua}). If
the ATIC and the observed hemispheric attenuation anomaly are
explained by solidification texturing controlled by the mantle
\cite{Aubertetal2008}, a differential rotation faster than a full
revolution in a few hundreds of millions of years would erase any
longitudinal signature in the texturing
\cite{Dumberry2010,Deguen2012}.

In order to avoid this contradiction, the recent dynamo models
adjust the viscous, magnetic, or gravitational torques such that the
present day mean rotation rate appears as a temporal fluctuation
superposed over a very slow steady super rotation of a few degrees
per million years \cite{Dumberry2010,AubertandDumberry2011} or even
without offset of steady super rotation over long timescales
\cite{DumberryandMound2010}. However, this is yet not a satisfactory
solution, since the longitudinal variations of the solidification
rates interacting with mantle heterogeneities can result in a
gravitationally driven differential rotation of the IC in the
westward direction \cite{Dumberry2010}, opposite to that of the
rotation suggested by seismic observations, which could be too fast
to preserve longitudinal differences in texturing \cite{Deguen2012}.

Another tentative explanation for ATIC and the attenuation anomaly
is a convective west-east translation of the IC
\cite{Alboussiereetal2010,Monnereauetal2010}, with minimal influence
of the mantle on the core dynamics \cite{Deguen2012,Tkalcicetal2013}
(see Appendix~\ref{apphist}). This theory considers that the IC is
shifted by 100~m from the Earth's center. But then, the viscous
momentum due to the OC flow near the asymmetric ICB should modify
the rate and the axis of the differential rotation, hindering the
onset of the internally driven convective translation. An additional
difficulty of this approach is the fact that flows derived from
geomagnetic observations and dynamo simulations reproducing patterns
of geomagnetic secular variation predict opposite IC translations
from east to west \cite{Aubert2013,Aubertetal2013}.

The decentered IC hypothesis avoids these contradictions because
both the ATIC and the hemispheric asymmetry of the attenuation are
explained by the displacement of the IC, without any constrains on
its internal structure (Sects.~\ref{resid} and \ref{atenua}). Then
the IC can rotate with respect to the mantle with any angular
velocity around any axis. Therefore our hypothesis does not rule out
any of the different existing estimations of the angular velocity of
the differential rotation which are listed below.

Early geodynamo predictions of a super-rotation of the IC by
$2-3^{\circ}$ per year induced by magnetic and viscous torques
exerted by the eastward zonal flow near the IC
\cite{GlatzmaierandRoberts1996} were found in reasonable agreement
with seismic data. Travel time anomalies of the PKIKP phase
indicated a differential rotation rate of $\sim 3^{\circ}$/yr
\cite{Suetal1996} while travel time residuals of the PKIKP rays with
respect to seismic rays turning at the base of the OC (PKPbc)
indicated a rate of $\sim 1^{\circ}$/yr \cite{SongandRichards1996}.
Subsequent studies, using earthquake doublets, normal modes, and
refined methods favor much smaller rotation rates between
$0.0^{\circ}$/yr and $0.3^{\circ}$/yr or even a retrograde,
westward, rotation \cite{Souriau2007}. Observations of temporal
trends of PKPbc-PKIKP residuals, similar to previous studies
\cite{SongandRichards1996} but using earthquakes produced in four
different source regions, were found to be consistent with both
eastward and westward differential IC rotation in different regions,
suggesting that differential IC rotations of tenths of degrees per
year are incompatible with global data \cite{MakinenandDeuss2011}.
On the other hand, an inverse analysis of doublets observed at
College station over more than forty years \cite{Tkalcicetal2013}
found an average differential rotation of $0.25-0.48^{\circ}$/yr and
decadal fluctuations of the order of $1^{\circ}$/yr. These results
are consistent with the observed decadal changes in the length of
the day \cite{Tkalcicetal2013}, which in turn indicate the presence
of zonal flows in the fluid core, as required by the currently
accepted mechanism of the IC differential rotation
\cite{GlatzmaierandRoberts1996,Deguen2012}, as well as with decadal
changes in the magnetic field which may be responsible for the
observed fluctuations in the IC rotation \cite{Livermoreetal2013}.

\subsection{Large-scale anomalies of the geoid\label{geoid}}

A displacement of the IC over tens of kilometers should produce a
measurable effect on the gravitational field at the Earth's surface.
A simple calculation shows that for an IC displacement of 1~km, the
surface of the geoid (the gravity equipotential that coincides with
sea level) changes by 1~m. The largest negative geoid anomaly has
-103 m and lies south of India while the largest positive one has 80
m and occurs near New Guinea. Hence the magnitude of the observed
anomalies corresponds to a displacement of about 100~km, but their
distribution has no explicit correlation with the theoretical
gravitational field of a displaced IC. At first sight, the observed
anomalies lack the necessary rotational symmetry around the
displacement direction of the IC determined in the previous
sections.

However, Barta \cite{barta1973} showed that the geoid anomalies can
be well approximated at large-scale by a superposition of two global
anomalies with cylindrical symmetries around axis directed toward
the centers of the two greatest geoid anomalies
\begin{widetext}
\begin{eqnarray*}
\Sigma_1(\theta,\phi)&=
&-\ \ 0.3-\ 3.70\,P_2(\theta_1)-46.11\,P_3(\theta_1)+\ \,8.16\,P_4(\theta_1)+15.56\,P_5(\theta_1) \\
\Sigma_2(\theta,\phi)&=
&-11.8+61.74\,P_2(\theta_2)+43.50\,P_3(\theta_2)-25.13\,P_4(\theta_2)-\
\,5.10\,P_5(\theta_2) ,
\end{eqnarray*}
\end{widetext}
where $P_i$ are the Legendre polynomials of order $i$. The angles
$\theta_1$ and $\theta_2$ are the angles between the radius with the
geographical coordinates $(\theta,\phi)$ and the axis pointing to
the directions of $58\,^{\circ}\mathrm{E}$ and
$156.5\,^{\circ}\mathrm{E}$ in the equatorial plane, respectively.
In this way, instead of several anomalies, we have to explain only
two global anomalies which already have the necessary cylindrical
symmetry. Barta ascribed $\Sigma_2$ to an eastward shift of the IC
over 100~km \cite{barta1981}. But the gravitational anomaly
generated by the shifted IC alone is dominated by the term
containing $P_1(\theta_1)=cos \theta_1$ which is absent in the
formula for $\Sigma_2$.

The explanation of the geoid anomalies $\Sigma_1$ and $\Sigma_2$
could be obtained by the mechanism of the dynamic topography. This
is the usual method to derive the longest wavelength components of
the residual geoid \cite{hager1985}. The seismically-inferred
density contrasts in the lower mantle are negatively correlated with
the observed geoid. Therefore one needs a counteracting effect,
called dynamic topography: the thermally driven convective flow in
the mantle induces deformations of Earth's surface and of the CMB
which count as negative masses \cite{pekeris1935}. The strength of
such a mechanism generating large-scale density variations inside
the OC is however disputed \cite{bowin2000}.

Theoretical estimations indicate that dynamical forces originating
inside the OC cannot support any appreciable internal structure
\cite{Stevenson1987}. But this is not true if there exists a
non-hydrostatic forcing due to gravitational interactions with the
exterior mass distributions or to boundary effects \cite{Wahr1989}.
As we pointed out in Sect.~\ref{nonhydro}, such a non-hydrostatic
forcing could be generated by the decentered IC. The results of
seismological investigations are contradictory. Some studies exclude
any large-scale aspherical structure inside the OC
\cite{Boschi2000,Ishii2005,Leykam2010,Souriau2003}, while others
have found significant cylindrical structures related to the tangent
cylinder (see Sect.~\ref{flow})
\cite{RomanowiczandBreger2000,Romanowicz2003} or non-homogeneous
structures inside the OC \cite{Dai2008,Piersanti2001,Soldati2003}.

Hence, the eastward displacement of the IC could cause a temperature
increase in the Eastern Hemisphere and implicitly a density
decrease. In addition, it could induce large-scale inhomogeneities
in the mantle structure and variations of the CMB depth. These
hemispheric asymmetries could compensate the IC displacement
explaining the absence of the term $P_1(\theta_1)$ in $\Sigma_2$.
This interpretation is in accordance with Bowin's estimation that
the second order term, which dominates the expression of $\Sigma_2$,
has the origin at depths of 6000~km, i.e., inside the IC
\cite[Fig.~8]{bowin2000}). The third degree term, dominating the
expression of $\Sigma_1$, originates at 3000~km depth, in the
neighborhood of CMB. In this way the large-scale anomalies of the
geoid are explained by two global density anomalies, one determined
by an eccentric IC, the other related to the topography of the CMB
and corresponding to the boundary conditions used in the numerical
simulation of the geodynamo \cite{Aubertetal2013}, discussed in
Sect~\ref{flow}.

\subsection{Hemispherical asymmetry of the IC anisotropy\label{anisotropy}}

The elastic anisotropy of the IC is characterized by P-waves travel
times about 3\% faster in the polar direction than parallel to the
equatorial plane. According to many seismic studies, the first
150~km beneath the ICB have low levels of anisotropy (smaller than
1\%) and the uppermost 50-90~km form an isotropic layer
\cite{Souriau2007,IrvingandDeuss2011} (see Sect.~\ref{resid}).
Anisotropy has been clearly documented only at depths of several
hundreds of kilometers
\cite{Souriau2007,NiuandChen2008,IrvingandDeuss2011}. While the
anisotropy in the innermost IC presented no significant longitudinal
dependence \cite{IshiiandDziewonski2002,NiuandChen2008},
hemispherical patterns were observed at depths above 600-700 km, with
an anisotropic Western Hemisphere and a nearly isotropic eastern one
\cite{Deussetal2010}, with anisotropy levels of 4.4\% and 1\%,
respectively \cite{IrvingandDeuss2011}. Recently, based on an
extensive data set of PKIKP travel times, Lythgoe et al. conclude
that there is no clear evidence of an innermost IC and the
hemispherical variation of the anisotropy extends through the entire
IC \cite{Lythgoe2014}.

There exist several theories regarding the generation of the IC
anisotropy with hemispherical variations \cite{SumitaandBergman2007}.
If it results from hemispheric variations in the IC growth, then a
mechanism is needed to produce different solidification rates and
crystalline structures inside the IC. This mechanism could be controlled by
interactions of the IC with the rest of the Earth, for instance, by
the OC flow caused by thermal inhomogeneities in the mantle
\cite{SumitaandBergman2007}. To ensure a constant influence on the
IC growth through such mechanisms, the IC should be locked to the
mantle for at least 200 million years \cite{IrvingandDeuss2011}, at
variance with the currently accepted theories on differential
rotation (see Sect.~\ref{diffrot}).

The hemispherical asymmetry in anisotropy of the IC is the mostly
used explanation for the complexity of the seismological
observations, but not the only one available. For example,
Romanowicz and Br\'{e}ger showed that anomalous splitting of most of
the normal modes sensitive to the interior structure of the IC,
excepting a single one, can be explained either by considering an
asymmetric IC anisotropy or an inhomogeneous OC structure
\cite{RomanowiczandBreger2000}. Indeed, polar regions of the OC with
seismic velocities larger than the velocities near the equatorial
plane generate an anisotropy of PKIKP travel times similar to that
generated by an anisotropic structure of the IC. Such an
inhomogeneous OC is consistent with the simulated flow inside the OC
(Sect.~\ref{flow}), which has different characteristics outside the
tangent cylinder and near the rotation axis, as well as with the
density inhomogeneities in the OC discussed in Sect.~\ref{geoid}.
Hence a decentred IC could be a part of a complete explanation of
the hemispherical asymmetry of the IC anisotropy.

\section{Conclusions}

An IC displaced by tens of kilometers from the Earth's center should
influence many mechanical, thermal, and magnetic phenomena in the
Earth's interior. In this paper we have analyzed several of them:
\begin{itemize}
    \item secular westward drift of the geomagnetic field at low
     latitudes (Sect.~\ref{flow});
    \item global asymmetries of the flow and magnetic field inside
     the OC (Sect.~\ref{flow});
    \item nonhydrostatic shape of the Earth (Sect.~\ref{nonhydro});
    \item Markowitz wobble of the Earth's pole
     (Sect.~\ref{nonhydro});
    \item decadal variations of the length of day
     (Sect.~\ref{nonhydro});
    \item translational oscillations of the IC (Slichter mode)
     (Sect.~\ref{slichter});
    \item isotropic hemispheric asymmetry of the travel times at the top
     of the IC (Sect.~\ref{resid});
    \item hemispheric asymmetry of the attenuation at the top of the IC
     (Sect.~\ref{atenua});
    \item hemispheric asymmetry of the precritical PKiKP travel
     times (Sect.~\ref{precrit});
    \item anomalous layer at the bottom of the OC (Sect.~\ref{anomal});
    \item differential rotation of the IC (Sect.~\ref{diffrot});
    \item large-scale anomalies of the geoid (Sect.~\ref{geoid});
    \item density heterogeneities inside the OC (Sect.~\ref{geoid});
    \item anisotropy of the IC (Sect.~\ref{anisotropy}).
\end{itemize}

We have approached most of the above issues qualitatively and
explained two of them (ATIC and the attenuation asymmetry) through
rigorous numerical simulations of differential travel times sampling
the top of the IC. In conclusion, we estimate that the displacement
of the IC could be larger than 3~km and smaller than 100~km, most
likely equal to several tens of kilometers. The estimated direction
of the displacement varies between $110\,^{\circ}\mathrm{E}$ and
$160\,^{\circ}\mathrm{E}$. The time scale of the translational
motion of the IC could be larger than several hundreds of years. The
resulting picture is that of a perturbation generated by the
decentered IC which enhances the nonequilibrium and the
nonhydrostaticity of the well-established basic structure of the
Earth. We consider that more flexible dynamical models of the
Earth's interior could be obtained by simply letting the IC to
participate in the movement of the OC.

It is tempting to adopt such a simple solution for a wide variety of
disputed geophysical problems. But for each individual phenomenon
there are many other perturbations making it difficult to obtain a
clear separation of the effects of a decentered IC. Nevertheless, we
believe that the numerous implications discussed in this article are
convincing enough and that this hypothesis deserves further exploration.

\appendix

\section{History of the eccentric IC hypothesis \label{apphist}}

The hypothesis of an eccentric IC with respect to the mass center of
the Earth has been launched for the first time in 1970s. At that time
the seismical observations were not providing enough
information on the Earth's structure near the ICB.
Therefore indirect observations regarding the gravity and geomagnetic fields
or the distribution of the crust inhomogeneities were used.

Barta interpreted the displacement of the dipole of the geomagnetic
field toward Australia as an indication for an eccentric IC
\cite{barta1973,Barta1974,barta1981}. Assuming that the density of
the OC is larger in the displacement direction, he found that the
distance between the IC and the Earth's rotation axis is of the
order of 100~km. He discussed the implications of such a
configuration on the magnetic secular variation, on the connection
between the magnetic and gravity fields, and on the deviation of
the geoid figure from the hydrostatic equilibrium.

From paleomagnetic data, Zidarov reached the conclusion that ``the
optimal magnetic dipole was located away from the Earth's center
during the geological past'' and, in consequence, ``the core itself
was located in the geological past away from the Earth's center,
toward the middle of the Pacific Ocean and gradually shifted toward
its center'' \cite{Zidarov1977}. Using the difference between the
equatorial moments of inertia of the Earth, he computed a
displacement of the IC of 35~km. He also discussed several other
effects of an eccentric IC on the piriform shape of the Earth, the
opening rate of the oceans, the geographic pole wandering, the
contemporary geotectonic activity, and the continental drift.

Vesanen and Teisseyre analyzed the deviations from symmetry of the
earthquake zones and interpreted them as an
indication for the existence of some deep asymmetry inside the Earth
\cite{Vesanen1978}. They also applied Barta and Zidarov's approaches
to assess the influence of an eccentric IC on the pattern of
convection cells in the Earth's interior.

A recently proposed mechanism which explains the hemispheric ATIC
(see Sect.~\ref{resid}) assumes an eastward translation of the IC as
a result of crystallization in the Western Hemisphere and melting in
the Eastern Hemisphere in a superadiabatic regime
\cite{Alboussiereetal2010,AlboussiereandDeguen2012,MizzonandMonnereau2013,Monnereauetal2010}.
In this process, the IC center of mass is shifted toward its colder
and denser Western Hemisphere \cite{Monnereauetal2010}. In an
attempt to restore the mechanical equilibrium, the IC as a whole
will then be shifted toward east. The eastward shift of the IC was
estimated at about 100~m in the equatorial plane
\cite{Alboussiereetal2010}. This shift produces a positive
topography on the Eastern Hemisphere of the ICB which melts by
exchange of latent heat with the OC fluid and, at the same time,
forces crystallization on the western IC hemisphere. A continuous
translation mechanism could result from the interaction of
superadiabaticity, gravitational equilibrium, and latent heat
exchange with the OC \cite{AlboussiereandDeguen2012}. The increase
of the iron grain-size during the west-east convective translation
of the IC could explain the observed hemispheric asymmetry of travel
times and attenuation.

Currently accepted parameters of the Earth's interior (e.g.
viscosity, IC age) impose constraints on the onset of this mechanism
\cite{Deguen2012} and it is currently difficult to reach firm
conclusions on the possibility of convection in the actual IC
\cite{DeguenandCardin2011}. Nevertheless, it is possible that
conditions for convection were met at early stages generating the
seismical asymmetry and anisotropy observed in the deeper parts of
the IC \cite{Buffett2009,Cormier2007}. In addition, the results of
recent geodynamo modeling of the geomagnetic secular variation are
rather consistent with a translation of the IC in the opposite,
east-west direction \cite{Aubert2013,Aubertetal2013}. Another
difficulty encountered by the proposed continuous convection
mechanism is the assumption that the IC is in mechanical equilibrium
with the surrounding fluid OC. The interaction of the flow at the
bottom of the OC with the positive topography of the ICB would
induce a supplementary rotation of the IC, at variance with the
static equilibrium and the fixed direction of the convective
translation predicted by the theory.

\section{Numerical model\label{appnumer}}

Since we focus only on the effect of a decentered inner core on the
propagation of the seismic rays, we consider a numerical model as
simple as possible so that other perturbing effects are disregarded.
Therefore, we keep most of the spherical symmetry of the Earth's
structure, even in case of a displaced inner core. The surfaces of
the Earth and of the inner core are spheres and the velocity profile
inside them is that of the one-dimensional model ak135
\cite{Kennettetal1995}. Outside the decentered inner core, the model
ak135 is linearly extrapolated to the points of the outer core
situated at distances from the Earth's center smaller than the
inner core radius. In this way we alter as little as possible the
model with spherical symmetry of the Earth's interior, maintaining
the symmetry separately for the inner core and the rest of the
Earth.

The inner core interior and the rest of the Earth are divided into
spherical layers with constant velocity of maximum 1~km thickness.
The reference levels of the ak135 model are all included among the
boundaries of spherical layers. Therefore the numerical seismic rays
are made of straight segments satisfying the refraction and
reflection laws at the boundaries of the spherical layers. The
propagation plane of the seismic ray changes only at the incidence
with the inner core boundary, which separates the two volumes with
spherical symmetry into the interior and exterior of the inner core.
The numerical errors of the travel time for a centered inner core
obtained with this numerical algorithm with respect to the values given
in seismological tables \cite{Kennett2005} are of the order of 0.01~s, i.e.,
one order of magnitude smaller than the residuals characterizing the
ATIC (Sect.~\ref{resid}).

\begin{figure}
\includegraphics{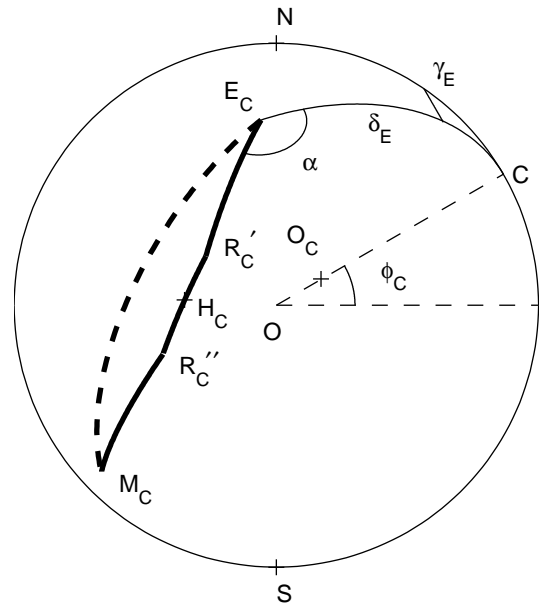}
\caption{\label{doubleproj}Double projection (on the Earth's surface
and on the meridian plane containing the center of the decentered
inner core) of the PKiKP$_{\mathrm{dec}}$ seismic ray. The dashed
line is the double projection of the PKIKP ray with the same focus
and exit points in case of the centered inner core. The meaning of the
points and angles notations are given in text.}
\end{figure}

\begin{figure}
\includegraphics{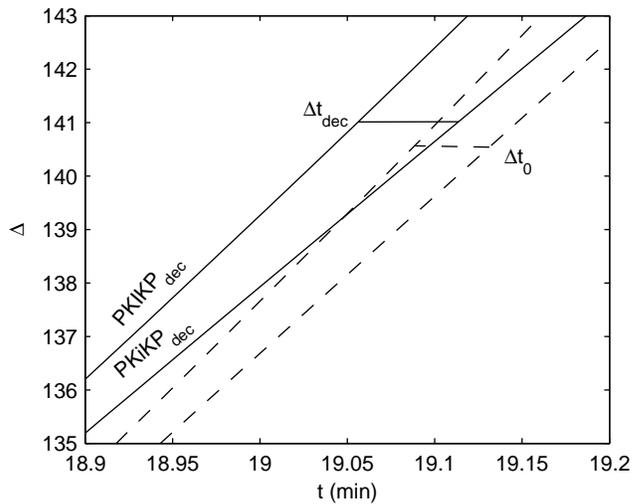}
\caption{\label{diagr}Seismic diagram for a displaced inner core.
Epicentral distance versus travel time for the
PKIKP$_{\mathrm{dec}}$ and PKiKP$_{\mathrm{dec}}$ rays in the
Eastern Hemisphere plotted in Fig.~\ref{asim} (continuous lines) and
for the PKIKP and PKiKP rays corresponding to a centered inner core
(dashed lines).}
\end{figure}

\begin{figure*}
\includegraphics{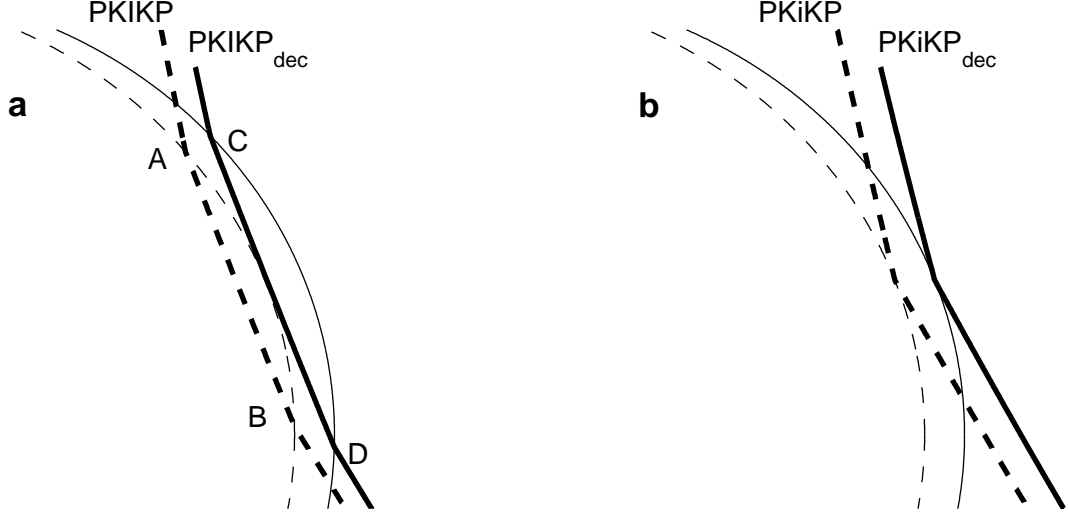}
\caption{\label{icbray}Seismic rays in the neighborhood of the inner
core boundary. The seismic rays have the same focus and exit point
and propagate in the meridian plane containing the center of the
inner core displaced in the Eastern Hemisphere as in Fig.~1. The
refracted rays (a) and the reflected ones (b) are plotted with thick
continuous lines for the decentered inner core and with thick dashed
lines for the centered inner core. The circular arcs represent boundaries
of the decentered (continuous line) and centered (dashed line) inner core.}
\end{figure*}

In order to construct the seismic ray we have to specify the
position of the earthquake focus and the initial propagation
direction of the ray. We first fix the latitude $\phi_{\mathrm{C}}$
and the longitude $\lambda_{\mathrm{C}}$ of the point C defined by
the intersection with the Earth's surface of the
straight line through the center O of the Earth and the center
O$_{\mathrm{C}}$ of the displaced inner core intersects the Earth's
surface (Fig.~\ref{doubleproj}). To avoid an intricate
three-dimensional graphical representation we use a double projection: first a
central projection on the Earth's surface from its center O and then
an orthogonal projection on the meridian plane containing the axis
OC defined by the displacement of the inner core.

The epicenter E is the projection of the earthquake focus on the
Earth's surface and E$_{\mathrm{C}}$ in Fig.~\ref{doubleproj} is the
projection of E on the meridian plane. The position of the epicenter
E in the plane OEC is given by the angle $\delta_{\mathrm{E}}$
between the radii OE and OC. The angle between the OEC plane and the
meridian plane NCS is denoted by $\gamma_{\mathrm{E}}$. The
propagation of the seismic ray has a cylindrical symmetry with
respect to the OC axis, its shape being independent of the angle
$\gamma_{\mathrm{E}}$.

Because of the spherical symmetry of the Earth's interior above
the inner core boundary, the seismic ray keeps its propagation plane
until it reaches the inner core boundary at the point R$^{'}$. The
propagation plane changes under reflection and refraction at the inner
core boundary. The PKIKP$_{\mathrm{dec}}$ phase propagates through
the displaced inner core with spherically symmetric internal
structure in the plane OR$^{'}$R$^{''}$ where R$^{''}$ is the exit
point on the inner core boundary. After the second refraction on the inner
core boundary, outside the inner core it propagates in the plane
OMR$^{''}$, where M is the exit point on the Earth's surface. The
PKiKP$_{\mathrm{dec}}$ seismic ray changes its propagation plane
once, at reflection on the inner core boundary at the point R$^{'}$.

In Fig.~\ref{doubleproj} the double projection
E$_{\mathrm{C}}$R$^{'}_{\mathrm{C}}$R$^{''}_{\mathrm{C}}$M$_{\mathrm{C}}$
of PKIKP$_{\mathrm{dec}}$ seismic ray is composed by three parts
separated by the deflection points where the seismic ray is
refracted by the inner core boundary. The initial propagation plane
OER$^{'}$ of the seismic ray is specified by the angle $\alpha$ with
the plane OEC. The initial direction of propagation in this plane is
given by the incidence angle between the ray and the normal to the
Earth's surface in E (not shown in the figure).

A special situation occurs when the seismic ray propagates in the
OEC plane ($\alpha=0$ or $\alpha=\pi$). Then, the normal to the inner
core boundary at R$^{'}$ belongs to the propagation plane, which
does not change by reflection or refraction at the inner core boundary.
The seismic rays for the same initial incident angle clearly exhibit
an east-west asymmetry, as shown in Fig.~\ref{asim} for an eastward
displacement of the inner core in the equatorial plane
($\lambda_{\mathrm{C}}=\pi/2$ and $\phi_{\mathrm{C}}=0$) and for
epicenter at the North Pole ($\delta_{\mathrm{E}}=\pi/2$).

When the inner core is decentered, the epicentral distance and the
travel time of the PKIKP$_{\mathrm{dec}}$ and PKiKP$_{\mathrm{dec}}$
phases depend on the initial propagation plane of the seismic ray
and on the location of the earthquake focus. First we analyze the
simple situation presented in Fig.~\ref{asim} when the seismic ray
is contained in the plane OCE and the inner core is shifted by
100~km toward 90$^{\circ}$~E ($\lambda_{\mathrm{C}}=\pi /2$ and
$\phi_{\mathrm{C}}=0$). For a focus depth of 200~km, we have
generated seismic rays with epicentral distance in the same interval
as that investigated in \cite{Waszeketal2011,WaszekandDeuss2011}.
The corresponding seismic diagram is presented in Fig.~\ref{diagr}.
We have chosen the PKIKP rays propagating in the region toward which
the inner core is shifted ($\alpha=0$), so that their path into the inner core is
longer than for a centered inner core. The diagrams for both
PKIKP$_{\mathrm{dec}}$ and PKiKP$_{\mathrm{dec}}$ phases are shifted
toward smaller travel times with a different amount, so that the
distance between diagrams changes and the residuals $\Delta
t_{\mathrm{dec}}-\Delta t_0$ are non-vanishing.

To explain the changes in seismic diagrams caused by the
displacement of the inner core, we plot in Fig.~\ref{icbray} the
paths in the neighborhood of the inner core boundary of the seismic rays
which emerge at the same point on the Earth's surface at $\Delta
=140^{\circ}$. Because of the shifted position of the inner core,
the total lengths, from the focus to the common exit point, of the
PKIKP$_{\mathrm{dec}}$ and PKiKP$_{\mathrm{dec}}$ rays are smaller
than those for a centered inner core (dashed lines) with
approximately the same amount. These shorter paths explain the
smaller travel times of the seismic phases for a decentered inner
core, but not the non-vanishing residuals $\Delta
t_{\mathrm{dec}}-\Delta t_0$.

There is another geometric effect which modifies the differential
travel time $\Delta t_{\mathrm{dec}}$. The segment CD of the seismic
ray within the decentered inner core is longer than the segment AB
for the centered inner core (Fig.~\ref{icbray}a). Because the
velocity in the inner core is larger than in the outer core, the
travel time of the PKIKP$_{\mathrm{dec}}$ phase has an additional
decrease, the differential travel time $\Delta t_{\mathrm{dec}}$
increases, and the residual $\Delta t_{\mathrm{dec}}-\Delta t_0$
becomes positive. In the diametrically opposite region of the inner
core the distance CD is smaller than AB and the residual is
negative, resulting in a hemispheric asymmetry. Hence, the asymmetry
of the residuals of the differential travel time can be explained by
the variation of the PKIKP$_{\mathrm{dec}}$ ray paths in the
decentered inner core without modifying the seismic velocities.

In order to obtain the geographical distribution of the ATIC
residuals plotted in Fig.~\ref{reprez} we compute them for
earthquakes evenly distributed on the Earth's surface and seismic
rays uniformly distributed around the focus. More precisely, we
specify the position of the inner core by fixing the angles
$\lambda_{\mathrm{C}}$ and $\phi_{\mathrm{C}}$ and the distance
between the center of the inner core and the center of the Earth to
$d=100$ km. The angles $\gamma_{\mathrm{E}}$, $\delta_{\mathrm{E}}$,
and $\alpha$ are then varied by steps of 10$^{\circ}$. For given
values of $\delta_{\mathrm{E}}$ and $\alpha$, we construct the
seismic ray with turning point at the depth of 39 km and we
calculate the corresponding residual. Because of the cylindrical
symmetry the shape of the seismic ray does not depend on
$\gamma_{\mathrm{E}}$. For each seismic ray we determine the
geographical coordinates of the turning point H (H$_{\mathrm{C}}$ in
Fig.~\ref{doubleproj} is its double projection).


\begin{thebibliography}{99}

\bibitem{Alboussiereetal2010}T. Alboussiere, R. Deguen, M.
Melzani, \textit{Melting-induced stratification above the Earth's
inner core due to convective translation}, Nature 466, 744-747
(2010).

\bibitem{AlboussiereandDeguen2012}T. Alboussiere, R. Deguen,
\textit{Asymmetric dynamics of the inner core and impact on the
outer core}, J. Geodyn. 61, 172-182 (2012).

\bibitem{Aubertetal2008}J. Aubert, H. Amit, G. Hulot,
P. Olson, \textit{Thermochemical flows couple the Earth's inner core
growth to mantle heterogeneity}, Nature 454, 758-761 (2008).

\bibitem{Aubert2013}J. Aubert, \textit{Flow throughout the Earth's
core inverted from geomagnetic observations and numerical dynamo
models}, Geophys. J. Int. 192, 537-556 (2013).

\bibitem{AubertandDumberry2011}J. Aubert, D. Dumberry,
\textit{Steady and fluctuating inner core rotation in numerical
geodynamo models}, Geophys. J. Int. 184, 162-170 (2011).

\bibitem{Aubertetal2013}J. Aubert, C.C. Finlay, A. Fournier,
\textit{Bottom-up control of geomagnetic secular variation by the
Earth's inner core}, Nature 502, 219-223 (2013).

\bibitem{barta1973}G. Barta, \textit{Physical background of the
geoidal figure}, Nature 243, 156 (1973).

\bibitem{Barta1974}G. Barta, \textit{Satellite geodesy and the
internal structure of the earth}, in \textit{Space research XIV},
Eds. M.J. Rycroft and R.D. Reasenberg, Akademie-Verlag, Berlin
(1974).

\bibitem{barta1981}G. Barta, \textit{Recent results of the study of physical
 background of the geoidal figure}, Adv. Space Res. 1, 195 (1981).

\bibitem{Boschi2000}L. Boschi, A.M. Dziewonski, \textit{Whole earth tomography from
delay times of P, PcP, and PKP phases: Lateral heterogeneities in
the outer core or radial anisotropy in the mantle?}, J. Geophys.
Res. 105, 13675-13696 (2000).

\bibitem{bowin2000}C. Bowin, \textit{Mass anomaly structure of the earth},
Review of Geophysics 38, 355-387 (2000).

\bibitem{Buffett2009}B.A. Buffett, \textit{Onset and orientation of
convection in the inner core}, Geophys. J. Int. 179, 711-719 (2009).

\bibitem{BuffettandBloxham2000}B. A. Buffett, J. Bloxham,
\textit{Deformation of Earth's inner core by electromagnetic
forces}, Geophys. Res. Lett. 27, 4001-4004 (2000).

\bibitem{BuffettandGlatzmaier2000}B. A. Buffett, G. A.
Glatzmaier, \textit{Gravitational braking of inner core rotation in
geodynamo simulations}, Geophys. Res. Lett. 27, 3125-3128 (2000).

\bibitem{CaoandRomanowicz2004}A. Cao, B. Romanowicz,
\textit{Hemispherical transition of seismic attenuation at the top
of the Earth's inner core}, Earth Planet. Sci. Lett. 228, 243-253
(2004).

\bibitem{christen2007}U.R. Christensen, J. Wiht, \textit{Numerical Dynamo
Simulations}, in \textit{Treatise on Geophysics}, Vol. 8,
\textit{Core Dynamics}, edited by G.
Schubert, P. Olson, 245-282, Elsevier,
Amsterdam (2007).

\bibitem{Cormier2007}
V. F. Cormier, \textit{Texture of the uppermost inner core from
forward- and back-scattered seismic waves}, Earth Planet. Sci. Lett.
258, 442-453 (2007).

\bibitem{Cormieretal2011}
V. F. Cormier, J. Attanayake, K. He, \textit{Inner core freezing and
melting: Constraints from seismic body waves}, Phys. Earth Planet.
Int. 188, 163-172 (2011).

\bibitem{Dai2008}W. Dai, X. Song, \textit{Detection of motion and heterogeneity in
Earth's liquid outer core}, Geophys. Res. Lett. 35, L16311 (2008).

\bibitem{Daietal2012}
Z. Dai, W. Wanga, L. Wen, \textit{Irregular topography at the
Earth's inner core boundary}, PNAS 109, 7654-7658 (2012).

\bibitem{Deguen2012}
R. Deguen, \textit{Structure and dynamics of Earth's inner core},
Earth Planet Sci. Lett. 333-334, 211-225 (2012).

\bibitem{DeguenandCardin2011}
R. Deguen, P. Cardin, \textit{Thermochemical convection in Earth's
inner core}, Geophys. J. Int. 187, 1101-1118 (2011).

\bibitem{dehant2007}V. Dehant, P.M. Mathews, \textit{Earth rotation
variations}, in \textit{Treatise on Geophysics}, Vol. 3,
\textit{Geodesy}, edited by G. Schubert, T.A. Herring, 239-294,
Elsevier, Amsterdam (2007).

\bibitem{Deussetal2010}A. Deuss, J.C.E. Irving, J. H.
Woodhouse, \textit{Regional variation of inner core anisotropy from
seismic normal mode observations}, Science 328, 1018-1020 (2010).

\bibitem{Dumberry2010grav}
M. Dumberry, \textit{Gravity variations induced by core flows},
Geophys. J. Int. 180, 635--650 (2010).

\bibitem{Dumberry2010}
M. Dumberry, \textit{Gravitationally driven inner core differential
rotation}, Earth Planet. Sci. Lett. 297, 387-394 (2010).

\bibitem{DumberryandMound2010}D. Dumberry, J. Mound,
\textit{Inner core-mantle gravitational locking and the
super-rotation of the inner core}, Geophys. J. Int. 181, 806-817
(2010).

\bibitem{Engdahletal1998}
E.R. Engdahl, R. van der Hilst, R. Buland, \textit{Global
teleseismic earthquake relocation with improved travel times and
procedures for depth determination}, Bull. Seismol. Soc. Am. 88,
722-743 (1998).

\bibitem{Galletetal2009}
Y. Gallet, G. Hulot, A. Chulliat, A. Genevey, \textit{Geomagnetic
field hemispheric asymmetry and archeomagnetic jerks}, Earth Planet.
Sci. Lett. 284, 179-186 (2009).

\bibitem{GarciaandSouriau2000}R. Garcia, A. Souriau,
\textit{Inner core anisotropy and heterogeneity level}, Geophys.
Res. Lett. 27, 3121-3124 (2000).

\bibitem{GlatzmaierandRoberts1996} G.A. Glatzmaier, P. H. Roberts,
\textit{Rotation and magnetism of Earth's inner core}, Science 274,
1887-1891 (1996).

\bibitem{gross2007}R.S. Gross, \textit{Earth rotation variations - long
period}, in \textit{Treatise on Geophysics}, Vol. 3,
\textit{Geodesy}, edited by G. Schubert, T.A. Herring, 239-294,
Elsevier, Amsterdam (2007).

\bibitem{Gubbins1981}D. Gubbins, \textit{Rotation of the inner core},
J. Geophys. Res. 86, 11695-11699 (1981).

\bibitem{Gubbinsetal2011}
D. Gubbins, B. Sreenivasan, J. Mound, S. Rost, \textit{Melting of
the Earth's inner core}, Nature 473, 361-363 (2011).

\bibitem{gwin86}C.R. Gwinn, T.A. Herring, I.I. Shapiro, \textit{Geodesy
by radio interferometry: Studies of the forced nutations of the
Earth, 2. Interpretation}, J. Geophys. Res. 91, 4755-4765 (1986).

\bibitem{hager1985}B.H. Hager, R.W. Clayton, M.A. Richards, R.P. Comer, A.M.
Dziewonski, \textit{Lower mantle heterogeneity, dynamic topography
and the geoid}, Nature 313, 541 (1985).

\bibitem{holme2007}R. Holme, \textit{Large-scale flow in the core},
in \textit{Treatise on Geophysics}, Vol. 8, \textit{Core Dynamics},
edited by G. Schubert, P. Olson, 107-130, Elsevier, Amsterdam
(2007).

\bibitem{Hulotetal2002}G. Hulot, c. Eymin, B. Langlais, M. Mandea, N. Olsen,
\textit{Small scale structure of the geodynamo inferred from Oersted
and Magsat satellite data}, Nature 416, 620-623, (2002).

\bibitem{ISC2013}
International Seismological Centre, On-line Bulletin,
http://www.isc.ac.uk, Internatl. Seis. Cent., Thatcham, United
Kingdom (2013).

\bibitem{Iritanietal2010}R. Iritani, N. Takeuchi, H. Kawakatsu,
\textit{Seismic attenuation structure of the top half of the inner
core beneath the northeastern Pacific}, Geophys. Res. Lett. 37,
L19303 (2010).

\bibitem{IrvingandDeuss2011}J.C.E. Irving, A. Deuss, \textit{A
Hemispherical structure in inner core velocity anisotropy}, J.
Geophys. Res. 116, B04307 (2011).

\bibitem{IshiiandDziewonski2002}M. Ishii, A. M. Dziewo\'{n}ski,
\textit{The innermost inner core of the earth: Evidence for a change
in anisotropic behavior at the radius of about 300 km}, PNAS 99,
14026-14030 (2002).

\bibitem{Ishii2005}M. Ishii, A.M. Dziewonski, \textit{Constraints on the outer-core
tangent cylinder using normal-mode splitting measurements}, Geophys.
J. Int. 162, 787-792 (2005).

\bibitem{jackson1993}A. Jackson, \textit{Intense equatorial flux
splots on the surface of the earth's core}, Nature 424, 760–763
(2003).

\bibitem{jackson2007}A. Jackson, C.C. Finlay, \textit{Geomagnetic secular
variation and its application to the core}, in \textit{Treatise on
Geophysics}, Vol. 5, \textit{Geomagnetism}, edited by G. Schubert,
M. Kono, 147-193, Elsevier, Amsterdam (2007).

\bibitem{JiangandZhao2012}
G. Jiang, D. Dapeng Zhao, \textit{Observation of high-frequency
PKiKP in Japan: Insight into fine structure of inner core boundary},
J. Asian Earth Sci. 59, 167-184 (2012).

\bibitem{Kawakatsu2006}
H. Kawakatsu, \textit{Sharp and seismically transparent inner core
boundary region revealed by an entire network observation of
near-vertical PKiKP}, Earth Planets Space 58, 855-863 (2006).

\bibitem{Kennett2005}
B.L.N. Kennett, \textit{Seismological Tables: ak135}, The Australian
National University Canberra ACT 0200 Australia (2005).

\bibitem{Kennettetal1995}B. Kennett, E. Engdahl, and R. Buland,
\textit{Constraints on seismic velocities in the Earth from travel
times}, Geophys. J. Int. 122, 108-124 (1995).

\bibitem{Koperetal2003}
K.D. Koper, M.L. Pyle, J.M. Franks, \textit{Constraints on
aspherical core structure from PKiKP-PcP differential travel times},
J. Geophys. Res. 108, 2168 (2003).

\bibitem{Lehmann1936}I. Lehman, \textit{P'}, Publications du
Bureau Central S\'{e}ismologique International A 14, 87-115 (1936).

\bibitem{Leykam2010}D. Leykam, H. Tkal\u{c}i\'{c}, and A.M. Reading, \textit{Core
structure re-examined using new teleseismic data recorded in
Antarctica: evidence for, at most, weak cylindrical seismic
anisotropy in the inner core}, Geophys. J. Int. 180, 1329-1343
(2010).

\bibitem{LiandCormier2002}X. Li, V.F. Cormier,
\textit{Frequency-dependent seismic attenuation in the inner core 1.
A viscoelastic interpretation}, J. Geophys. Res. 107, 2361 (2002).

\bibitem{Livermoreetal2013}P. W. Livermore, R. Hollerbach, A.
Jackson, \textit{Electromagnetically driven westward drift and
inner-core superrotation in Earth's core}, PNAS 110.40, 15914-15918
(2013).

\bibitem{Lythgoe2014}K.H.Lythgoe, A. Deuss, J.F. Rudgea, J.A. Neufeld,
\textit{Earth's inner core: Innermost inner core or hemispherical
variations?}, Earth Planet. Sci. Lett. 385, 181-189 (2014).

\bibitem{Mandeaetal2012}
M. Mandea, I. Panet, V. Lesur, O. de Viron, M. Diament, J-L. Le
Mou\"{e}l, \textit{Recent changes of the Earth's core derived from
satellite observations of magnetic and gravity fields}, PNAS 109.47,
19129-19133 (2012).

\bibitem{MakinenandDeuss2011}A. M. M\"{a}kinen, A. Deuss,
\textit{Global seismic body-wave observations of temporal variations
in the Earth's inner core, and implications for its differential
rotation}, Geophys. J. Int. 187, 355-370 (2011).

\bibitem{Monnereauetal2010}M. Monnereau, M. Calvet, L. Margerin,
A. Souriau, \textit{Lopsided growth of Earth's inner core}, Science
328, 1014-1017 (2010).

\bibitem{MizzonandMonnereau2013}
H. Mizzon, M. Monnereau, \textit{Implication of the lopsided growth
for the viscosity of Earth's inner core}, Earth Planet. Sci. Lett.
361, 391-401 (2013).

\bibitem{NiuandChen2008}F. Niu, Q. Chen, \textit{Seismic evidence
for distinct anisotropy in the innermost inner core}, Nature
Geoscience 1, 692-696 (2008).

\bibitem{NiuandWen2001}F. Niu, L. Wen, \textit{Hemispherical
variations in seismic velocity at the top of the Earth's inner
core}, Nature 410, 1081-1084 (2001).

\bibitem{ohtaki2012}T. Ohtaki, S. Kaneshima, K. Kanjo, \textit{Seismic
structure near the inner core boundary in the south polar region},
J. Geophys. Res. 117, B03312 (2012).

\bibitem{Okal2009}E.A. Okal, S. Stein, \textit{Observations of
ultra-long period normal modes from the 2004 Sumatra–Andaman
earthquake}, Phys. Earth Planet. In. 175, 53–62 (2009).

\bibitem{olsen2007}N. Olsen, G. Hulot, T.J. Sabaka,
\textit{The Present Field}, in \textit{Treatise on Geophysics}, Vol.
5, \textit{Geomagnetism}, edited by G.
Schubert, M. Kono, 147-193, Elsevier,
Amsterdam (2007).

\bibitem{OlsonandDeguen2012}
P. Olson, R. Deguen, \textit{Eccentricity of the geomagnetic dipole
caused by lopsided inner core growth}, Nature Geoscience 5, 565-569
(2012).

\bibitem{OreshinandVinnik2004}S.I. Oreshin, L.P. Vinnik,
\textit{Heterogeneity and anisotropy of seismic attenuation in the
inner core}, Geophys. Res. Lett. 31, L02613 (2004).

\bibitem{OuzounisandCreager2001}A. Ouzounis, K. Creager,
\textit{Isotropy overlying anisotropy at the top of the inner core},
Geophys. Res. Lett. 28, 4331-4334 (2001).

\bibitem{pekeris1935}C.L. Pekeris, \textit{Thermal convection in
the interior of the Earth}, Mon. Not. R. Astron. Soc., suppl. 3,
343-367 (1935).

\bibitem{Pengetal2008}
Z. Peng, K.D. Koper, J.E. Vidale, F. Leyton, P. Shearer,
\textit{Inner-core fine-scale structure from scattered waves
recorded by LASA}, J. Geophys. Res. 113, B09312 (2008).

\bibitem{Piersanti2001}A. Piersanti, L. Boschi, A.M. Dziewonski, \textit{Estimating lateral
structure in the earth's outer core}, Geophys. Res. Lett., 28,
1659-1662 (2001).

\bibitem{PoupinetandKennett2004}
G. Poupinet, B.L.N. Kennett, \textit{On the observation of high
frequency PKiKP and its coda in Australia}, Phys. Earth Planet. Int.
146, 497–511 (2004).

\bibitem{RomanowiczandBreger2000}B. Romanowicz, L. Br\'{e}ger,
\textit{Anomalous splitting of free oscillations: A reevaluation of
possible interpretations}, J. Geophys. Res. 105, 21559-21578 (2000).

\bibitem{Romanowicz2003}B. Romanowicz, H. Tkal\u{c}i\'{c}, L. Br\'{e}ger, \textit{On the
origin of complexity in PKP travel time data}, in \textit{Core
Dynamics, Structure and Rotation}, Dehant V et al. (eds.), pp.
31–44, American Geophysical Union, Washington (2003).

\bibitem{rosat2004}S. Rosat, \textit{Time varying gravity in relation
with the Earth's intern dynamics: Contribution of superconducting
gravimeters}, Doctoral Thesis, Strasbourg University, 2004.

\bibitem{roult2010}G. Roult, J. Roch, E. Clevede, \textit{Observation of
split modes from the 26th December 2004 Sumatra-Andaman mega-event},
Physics of the Earth and Planetary Interiors 179, 45–59 (2010).

\bibitem{ShearerandMasters1990}
P. Shearer, G. Masters, \textit{The density and shear velocity
contrast at the inner core boundary}, Geophys. J. Int. 102, 491-498
(1990).

\bibitem{slichter61}L.B. Slichter, \textit{The fundamental free mode
of the Earth's inner core}, PNAS 47, 186–190 (1961).

\bibitem{Soldati2003}G. Soldati, L. Boschi, A. Piersanti, \textit{Outer core density
heterogeneity and the discrepancy between PKP and PcP travel time
observations}, Geophysical research letters 30, 1190 (2003).

\bibitem{SongandRichards1996}X. Song, P. G. Richards,
\textit{Seismological evidence for differential rotation of the
Earth'sinner core}, Nature 382, 221-224 (1996).

\bibitem{Souriau2003}A. Souriau, A. Teste, S. Chevrot, \textit{Is there any structure
inside the liquid outer core?}, Geophys. Res. Lett. 30, 1567 (2003).

\bibitem{Souriau2007}A. Souriau, \textit{Deep Earth structure -
the Earth's cores}, in \textit{Treatise on Geophysics}, Vol. 1,
\textit{Seismology and structure of the Earth}, edited by G.
Schubert, B. Romanowicz, A. Dziewonski, 655-693, Elsevier, Amsterdam
(2007).

\bibitem{Stevenson1987}D.J. Stevenson, \textit{Limits on lateral density and velocity
variations in the earth's outer core}, Geophys. J. Roy. Astr. S. 88,
311-319 (1987).

\bibitem{Suetal1996}W. Su, A. M. Dziewonski, R. Jeanloz,
\textit{Planet within a planet: Rotation of the Inner Core of
Earth}, Science 274, 1883-1887 (1996).

\bibitem{SumitaandOlson1999}I. Sumita, P. Olson,
\textit{A laboratory model for convection in Earth's core driven by
a thermally heterogeneous mantle}, Science, 286, 1547-1549 (1999).

\bibitem{SumitaandOlson2002}I. Sumita, P. Olson,
\textit{Rotating thermal convection experiments in a hemispherical
shell with heterogeneous boundary heat flux: implications for the
Earth's core}, J. Geophys. Res. 107, 2169 (2002).

\bibitem{SumitaandBergman2007}I. Sumita, M.I. Bergman,
\textit{Inner-Core Dynamics}, in \textit{Treatise on Geophysics},
Vol. 8, \textit{Core Dynamics}, edited by  G.
Schubert, P. Olson, 299-318,
Elsevier, Amsterdam (2007).

\bibitem{tanaka1997}S. Tanaka and H. Hamaguchi, \textit{Degree one
heterogeneity and hemispherical variation of anisotropy in the inner
core from PKP(BC)-PKP(DF) times}, J. Geophys. Res. 102, 2925-2938
(1997).

\bibitem{Tkalcicetal2013}H. Tkal\u{c}i\'{c}, M. Young, T. Bodin,
S. Ngo, M. Sambridge, \textit{The shuffling rotation of the Earth's
inner core revealed by earthquake doublets}, Nature Geoscience 6,
497-502 (2013).

\bibitem{VamosandSuciu2011}C. Vamo\c{s}, N. Suciu, \textit{Seismic
hemispheric asymmetry induced by Earth's inner core decentering},
arXiv:1111.1121v1 [physics.geo-ph] (2011).

\bibitem{Vesanen1978}E. Vesanen, R. Teisseyre, \textit{Symmetry
and asymmetry in geodynamics}, Geophysica 15, 147-170 (1978).

\bibitem{Wahr1989}J. Wahr, D. de Vries, \textit{The possibility of lateral structure
inside the core and its implications for nutation and earth tide
observations}, Geophys. J. Int. 99, 511-519 (1989).

\bibitem{Waszeketal2011}L. Waszek, J. Irving, A. Deuss,
\textit{Reconciling the hemispherical structure of Earth's inner core with
its super-rotation}, Nature Geoscience 4, 264-267 (2011).

\bibitem{WaszekandDeuss2011}
L. Waszek, A. Deuss, \textit{Distinct layering in the hemispherical
seismic velocity structure of Earth's upper inner core}, J. Geophys.
Res. 116, B12313 (2011).

\bibitem {WenandNiu2002}L. Wen, F. Niu, \textit{Seismic velocity
and attenuation structures in the top of the Earth's inner core}, J.
Geophys. Res. 107(B11), 2273 (2002).

\bibitem{Yuetal2005}W. Yu, L. Wen, F. Niu, \textit{Seismic
velocity structure in the Earth's outer core}, J. Geophys. Res. 110,
B02302 (2005).

\bibitem{YuandWen2006}W. Yu, L. Wen, \textit{Seismic velocity
and attenuation structures in the top 400 km of the Earth's inner
core along equatorial paths}, J. Geophys. Res. 111, B07308 (2006).

\bibitem{Zidarov1977}D. Zidarov, \textit{Theory of the evolution of
the Earth and the Earth's crust based on the mobile Earth core
concept}, Geol. Balcanica 7, 3-26 (1977).

\end{thebibliography}
\end{document}